\documentclass{aa}  

\usepackage{graphicx}
\usepackage{txfonts}
\usepackage{siunitx}
\usepackage{comment}
\usepackage{caption}
\usepackage{subcaption}
\usepackage{graphics}
\usepackage{geometry}
\usepackage{hyphenat}
\usepackage{subcaption}
\usepackage{cleveref}
\usepackage{indentfirst}
\usepackage{gensymb}
\usepackage{newtxtext,newtxmath}
\usepackage[T1]{fontenc}
\usepackage{amsmath}
\usepackage{amssymb}

\makeatletter
\renewcommand*{\@fnsymbol}[1]{\ensuremath{\ifcase#1\or \star\or \dagger \else\@ctrerr\fi}}
\makeatother
 
\begin{document}

   \title{Photometric determination of the mass accretion rates of pre-main sequence stars. VII. The low-density cluster NGC 376 in the Small Magellanic Cloud\thanks{Based on observations with the NASA/ESA Hubble Space Telescope, obtained at the Space Telescope Science Institute, which is operated by AURA, Inc., under NASA contract NAS5-26555.}\thanks{Full Table 2 is only available at the CDS via anonymous ftp to cdsarc.cds.unistra.fr (130.79.128.5) or via https://cdsarc.cds.unistra.fr/cgi-bin/qcat?J/A+A/}}


   \author{Styliani Tsilia\inst{1}
          \and
          Guido De Marchi\inst{2}
          \and
          Nino Panagia\inst{3}
          }

   \institute{Sterrewacht Leiden, P.O. Box 9513, 2300 RA Leiden, the Netherlands\\
              \email{tsilia@strw.leidenuniv.nl}
         \and
             European Space Research and Technology Centre, Keplerlaan 1, 2200 AG Noordwijk, Netherlands\\
             \email{gdemarchi@esa.int}
        \and
            Space Telescope Science Institute, 3700 San Martin Drive, Baltimore, MD 21218, USA
             \email{panagia@stsci.edu}}

   \date{Received 5 December 2021; accepted 08 December 2022}

 
  \abstract
   {}
   {We study the properties of low-mass stars recently formed in the field of the NGC\,376 cluster in the Small Magellanic Cloud (SMC).}
   {Using photometric observations acquired with the Hubble Space Telescope (HST) in the $V$, $I$, and $H\alpha$ bands, we identify 244 candidate pre-main sequence (PMS) stars showing H$\alpha$ excess emission at the 5$\sigma$ level and with H$\alpha$ equivalent width of 20\,\AA\ or more. We derive physical parameters for all PMS stars, including masses, ages, and mass accretion rates. We compare the effective mass accretion rate of stars in NGC\,376 to that of objects in the NGC\,346 cluster, with NGC\,346 featuring similar metallicity but higher total mass and gas density.}
   {We find a median age of 28\,Myr for this population (with 25th and 75th percentiles at about 20 and 40\,Myr, respectively), in excellent agreement with previous studies of massive stars in the same field. The PMS stars are rather uniformly distributed across the field, whereas massive stars are more clustered. The spatial distribution of PMS objects is compatible with them having formed in the centre of the cluster and then migrating outwards. We find that in NGC\,376 the mass accretion rate is systematically lower than in NGC\,346 for stars of the same mass and age. This indicates that, in addition to metallicity, there are other environmental factors affecting the rate of mass accretion onto PMS stars. Our observations suggest that the gas density in the star-forming region might play a role.}
   {}

   \keywords{stars: formation -- stars: pre-main sequence -- galaxies: Magellanic Clouds -- galaxies: star clusters: individual (NGC\,376) -- open clusters and associations: individual (NGC\,376)
               }

   \maketitle
%

\section{Introduction}

\label{sec:intro}
The majority of stars in the Universe are low-mass objects. This is revealed by the shape of the initial mass function \citep[IMF; e.g.,][]{kroupa2001variation} of stellar populations: the peak of the mass distribution is found in the range $0.1-0.2$\,M$_{\odot}$. The process of low-mass star formation is therefore crucial to investigating the birth of most of the existing stars.

Our current understanding of low-mass star formation tells us that protostars form within the denser regions of large molecular clouds made up of gas and dust, when these regions collapse under their own gravitation from the inside out and in a self-similar manner \citep[e.g.][]{1977ApJ...214..488S}. These protostars are surrounded by circumstellar discs, and the young pre-main sequence (PMS) objects grow through the accretion of mass from these discs. Eventually the accretion is terminated, the circumstellar discs are dispersed, and stable systems emerge as the PMS objects reach the zero age main sequence (ZAMS). As a reference, a star like our Sun would need about 50\,Myr to reach the ZAMS \citep[e.g.][]{chen2015parsec}.

The rate of mass accretion is a fundamental property when studying the formation of low-mass stars; for instance, it has a direct impact on the final mass of these objects and hence on the functional shape of the IMF. Moreover, it affects the formation and migration of planets, by shaping the environment within which these occur. Although angular momentum transport by magnetic fields is speculated to be involved in the process of mass accretion onto PMS stars, the exact mechanism remains unclear \citep[see e.g.,][]{doi:10.1146/annurev-astro-081915-023347}.

It has been shown observationally that the rate of mass accretion of PMS stars depends on the mass and on the age of the forming object \citep{muzerolle2003accretion,muzerolle2005measuring,natta2004accretion,natta2006accretion,calvet2004mass,white2004evolutionary,sicilia2005cepheus}, although the uncertainties are relatively large, and \citet{clarke2006m} suggest that a very steep dependence might be spurious. It has been pointed out that mass and age are intertwined and cannot be regarded independently without causing biases when studying star formation processes unless, for example, a sample of PMS stars is split into smaller uniform groups of similar age \citep{Marchi_2017}.

Recently, metallicity was shown to also affect the intensity and duration of the mass accretion process, which brings into play the impact of the environment where star formation occurs. From a multi-parametric fit to the mass accretion rate of a uniform sample of thousands of PMS stars in the Milky Way and Magellanic Clouds, \citet{Marchi_2017} formulated the mass accretion equation as $\log\dot M_{acc}=a\times \log t +b\times \log m +c$, where $t$ is the age (in Myr) and $m$ the mass (in M$_{\odot}$). Their study showed that, while the values of $a$ and $b$ are common to all these regions (respectively  $a = -0.59 \pm 0.02$ and $b=0.78\pm0.08$), the $c$ parameter, which is the effective mass accretion rate per unit of measurement, appears to vary considerably across different regions, suggesting that there are extra environmental effects that must be considered. And indeed, \citet{Marchi_2017} discovered a correlation between $c$ and metallicity $Z$: for stars of the same mass and age, the mass accretion rate appears to be systematically higher at lower metallicity.

Nevertheless, the correlation between $c$ and $Z$ identified by these authors does show some scatter, suggesting that metallicity is not the only factor to influence mass accretion; other environmental conditions, for instance the gas density or the intensity of the local magnetic fields, might also contribute to the $c$ term in the previous equation. \citet{Marchi_2017} note the case of the Small Magellanic Cloud (SMC) cluster NGC\,602, which exhibits a great deviation from the best fit to the $c - Z$ relationship, which suggests the existence of other factors leading to lower mass accretion. Low gas density was considered a likely explanation in this situation since all other regions analysed in that work formed in denser environments than that of NGC\,602, which is located in the Wing of the SMC, where gas density is lower. A further exploration of the impact of gas density on star formation motivated this work, in which we study another low-density star-forming region in the SMC, namely NGC\,376.

This region, with a metallicity of about $1/5\,Z_{\odot}$, typical of the SMC \citep[see e.g.][]{russell1992abundances,rolleston1999chemical,lee2005chemical}, lies in the eastern extension (Wing) of the SMC, at a projected distance of $\sim 850$\,pc from the centre of the galaxy. The cluster is relatively young and loose. Age determinations from ground-based photometry, ranging from 16\,Myr \citep{chiosi2006age} to $\sim 25\pm10$\,Myr \citep{piatti2007young}, were confirmed by  \citet{Sabbi}, who used observations obtained with the {\em Hubble Space Telescope} (HST) in the $V$ and $I$ bands. By comparing the colour and magnitudes of massive stars with theoretical isochrones, they derived an age of $28\pm7$\,Myr. They also constructed surface brightness and radial density profiles that suggest that the cluster is not in virial equilibrium, but is actually in the process of dissolving.

We explicitly chose to study NGC\,376 because, thanks to its location in the Wing of the SMC, it combines low metallicity with low gas density of the environment, similar to NGC\,602. If metallicity were the only environmental factor affecting mass accretion rate, we would expect to find a value of $c$ for this cluster similar to that of massive SMC clusters such as NGC\,346. On the other hand, if its mass accretion rate turned out to be lower, similar to that of the loose NGC\,602 cluster, it would tell us that additional environmental effects such as density can play an important role in the star formation process.

This paper builds upon extensive previous work conducted on the Magellanic Clouds regarding star formation processes. As the environment of NGC\,376 differs considerably from that typical of Milky Way star-forming regions (primarily because of the metallicity), it is not surprising that differences in accretion and disc dissipation rates are observed. The ages of PMS stars in the Magellanic Clouds are also systematically larger; in small nearby associations with solar metallicity, there are very few actively accreting PMS stars older than 10\,Myr, while in massive star-forming regions in low-metallicity environments they appear to be much more numerous. 

The structure of this paper is as follows. In Section~\ref{sec:obs} we describe the observations that we used and how we analysed images from two different instruments on board the HST. In Section~\ref{sec:cmd} we present the photometry and the resulting colour--magnitude diagram (CMD), while in Section~\ref{sec:search} we identify candidate PMS stars and address their spatial distribution. In Section~\ref{sec:phys}, we derive and discuss the stellar physical parameters of these objects, including their mass accretion rates. A summary and conclusions follow in Section~\ref{sec:sum}.

\section{Observations and photometry}

\label{sec:obs}
\begin{figure}
    \centering
    \begin{subfigure}[b]{0.45\textwidth}
    \includegraphics[width=1\linewidth]{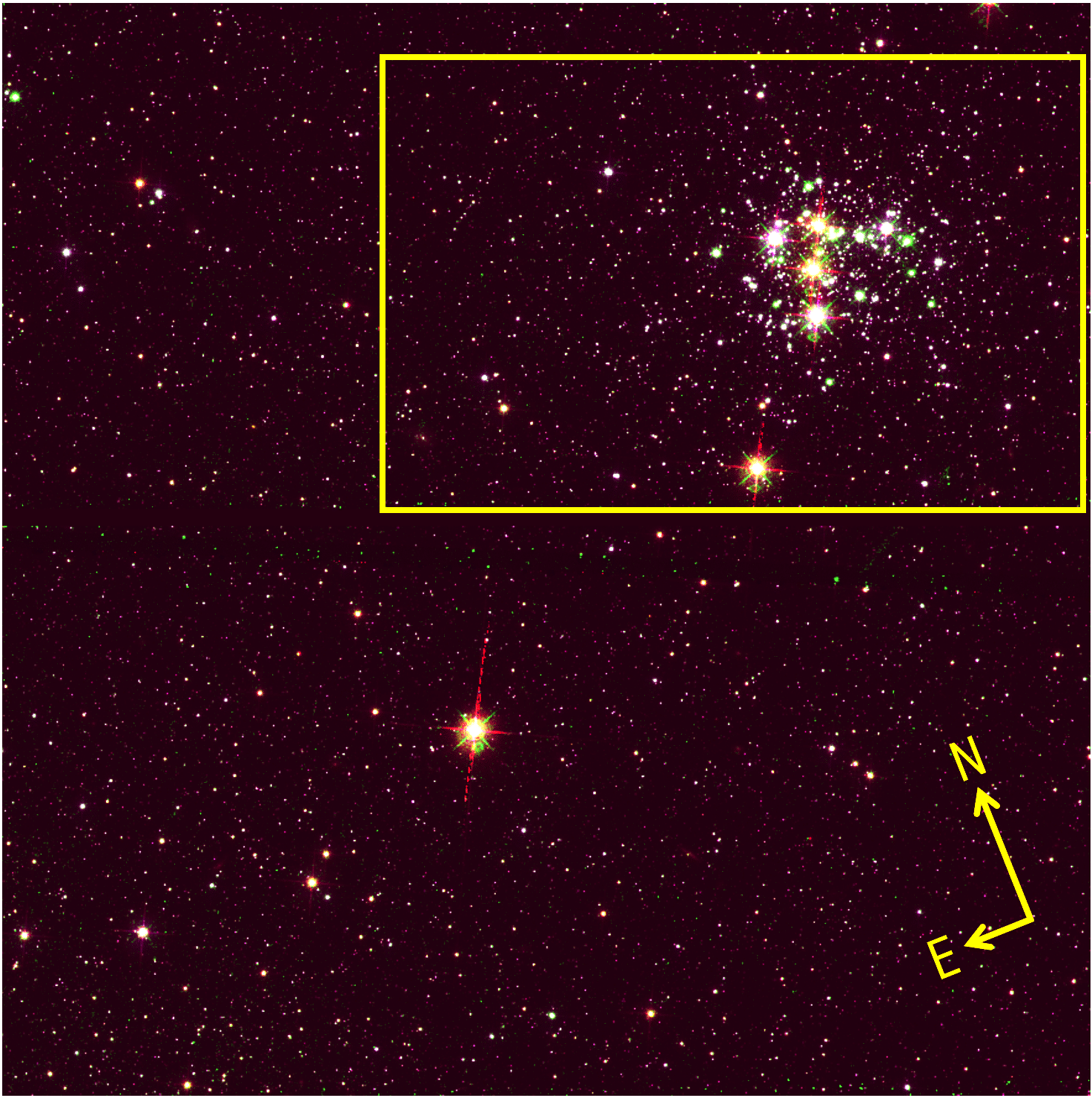}
    \caption{}
    \label{comp}
    \end{subfigure}

    \begin{subfigure}[b]{0.45\textwidth}
    \includegraphics[width=1\linewidth]{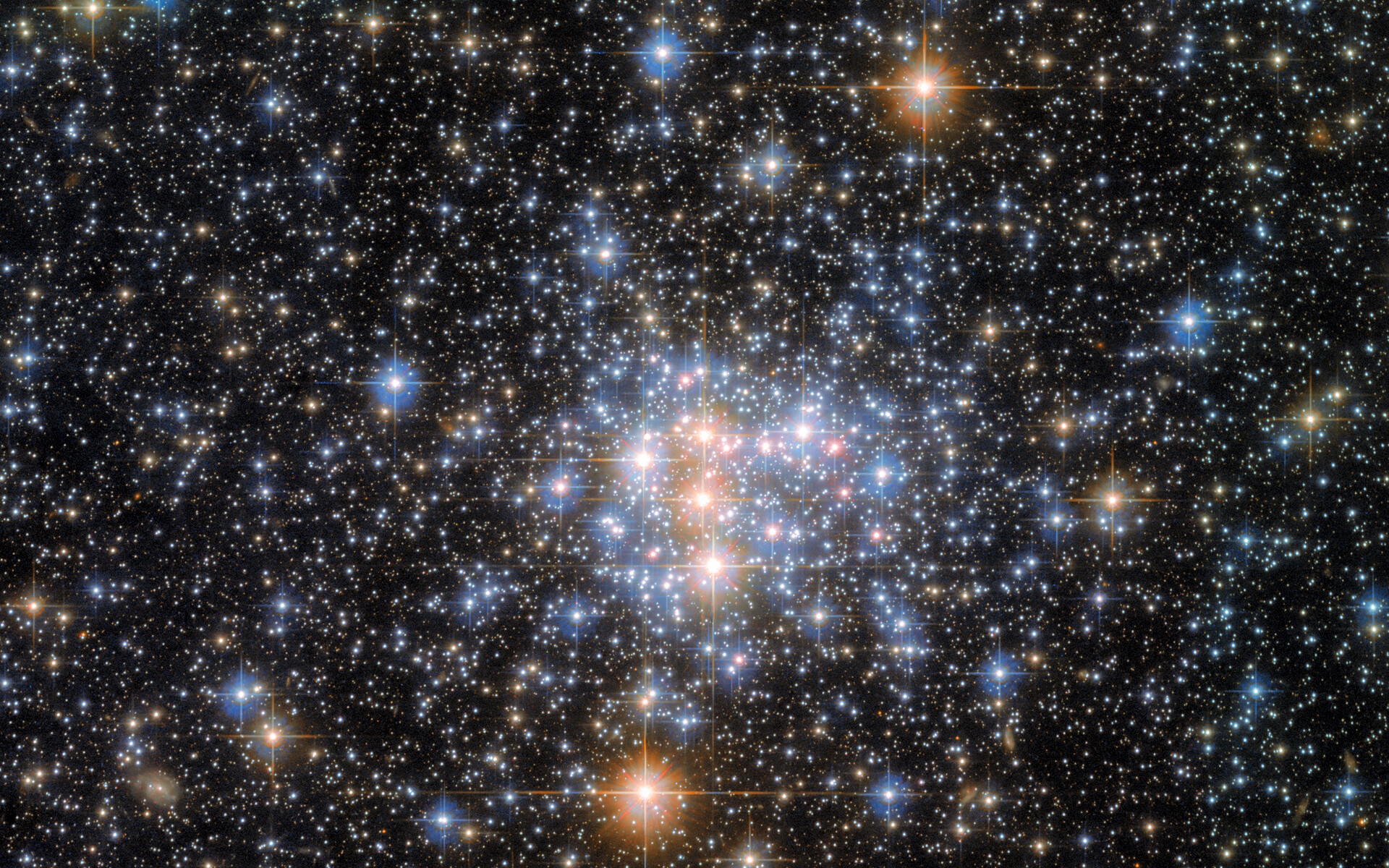}
    \caption{}
    \label{hubble}
    \end{subfigure}

    \caption{Low-density cluster NGC\,376 in the Small Magellanic Cloud. (a) Pseudo-colour image obtained by assigning the RGB channels to the $I$, $H\alpha$, and $V$ bands, respectively. The non-matching portions of the frame were masked. North is rotated 22 degrees anti-clockwise from the vertical. The image spans about 160\,arcsec on each side, which corresponds to a true size of approximately 48\,pc at the distance of the SMC. The yellow rectangle highlights the area depicted in (b). (b) Higher-resolution rendering of the innermost cluster regions (Credit: ESA/Hubble \& NASA, A. Nota, G. De Marchi).}
\end{figure}

We used images obtained with the {\em Wide Field Channel} (WFC) of the {\em Advanced Camera for Surveys} (ACS) and the  {\em UV/Visible} (UVIS) channel of the {\em Wide Field Camera 3} (WFC3) on board the HST. The ACS/WFC data were obtained in 2004 September (proposal number 10248, principal investigator: A. Nota) as part of a programme to study the properties of young star clusters in the Magellanic Clouds, and includes observations in the F555W band (hereafter $V$) and F814W band (hereafter $I$). Images were acquired in both of these filters, with the total exposure times being 1806\,s and 1966\,s in the $V$ and $I$ filters respectively. Additionally, and in order to correct for saturated stars, short exposures of 3\,s were used (one exposure for each dithering position and filter, amounting to four exposures in total). The WFC3/UVIS data were obtained in 2013 August (proposal number 13009, principal investigator: G. De Marchi) as part of another project to identify PMS stars in the Magellanic Clouds. This includes an image corresponding to a total exposure time of 2389\,s in the F656N band (hereafter $H\alpha$). A summary of the observations we used is given in Table~\ref{tab:obs}. A colour composite image of NGC\,376 is shown in Figure~\ref{comp}, combining ACS/WFC images in the $V$ and $I$ bands with the WFC3/UVIS image in the $H\alpha$ band.
\begin{table}
	\centering
	\caption{Journal of ACS and WFC3 Observations}
	\label{tab:obs}
	\resizebox{\columnwidth}{!}{%
	\begin{tabular}{cccccc} 
		\hline
		\textbf{Image Name} & \textbf{Date and Time of Observation} & \textbf{R.A.} & \textbf{Decl.} & \textbf{Filter} & \textbf{Exposure Time}\\
		\hline
		hst\_10248\_08\_acs\_wfc\_f555w & 12/09/04 13:52:45 & $01^{\rm h}04^{\rm m}06\mbox{$.\!\!^{\mathrm s}$}11$ & -72\degree49'57".1 & F555W & 1806.0\\
		hst\_10248\_08\_acs\_wfc\_f814w & 12/09/04 14:49:52 & $01^{\rm h}04^{\rm m}06\mbox{$.\!\!^{\mathrm s}$}11$ & -72\degree49'57".1 & F814W & 1966.0\\
		hst\_13009\_91\_wfc3\_uvis\_f656n & 01/08/13 20:01:06 & $01^{\rm h}04^{\rm m}04\mbox{$.\!\!^{\mathrm s}$}41$ & -72\degree49'51".9 & F656W & 2389.0\\
        J92F08LZQ & 12/09/04 13:52:45 & $01^{\rm h}03^{\rm m}53\mbox{$.\!\!^{\mathrm s}$}8$ & -72\degree49'29".6 & F555W & 3.0\\
        J92F08M9Q & 12/09/04 14:41:01 & $01^{\rm h}03^{\rm m}53\mbox{$.\!\!^{\mathrm s}$}8$ & -72\degree49'29".6 & F555W & 3.0\\
        J92F08MBQ & 12/09/04 14:49:52 & $01^{\rm h}03^{\rm m}53\mbox{$.\!\!^{\mathrm s}$}8$ & -72\degree49'29".6 & F814W & 3.0\\
        J92F08MLQ & 12/09/04 16:07:22 & $01^{\rm h}03^{\rm m}53\mbox{$.\!\!^{\mathrm s}$}8$ & -72\degree49'29".6 & F814W & 3.0\\
		\hline
	\end{tabular}%
    }
\end{table}

The photometric analysis was conducted with the help of the DAOPHOT routine in IRAF. We searched for unresolved objects with the count rate in their central pixel exceeding the median local background level by at least five times the local uncertainty. We then selected an aperture of 2 pixels in radius and a sky annulus from 3 to 5 pixels. An aperture correction was applied in order to account for the corresponding encircled energies, while stellar magnitudes were calibrated into the Vegamag photometric system by using a zero-point correction. Both of these corrections were implemented following the ACS and WFC3 data handbooks \citep{acsbook,wfc3book}.

A list of average photometric uncertainties as a function of magnitude in all three bands is given in Table~\ref{tab:errors}. The photometry in the $H\alpha$ band dominates the uncertainties, as expected, because of the lower throughput of the filter.

\begin{table}
	\centering
	\caption{Photometric catalogue of objects with uncertainty in H$\alpha$ smaller than $0.2$\,mag. The full photometric catalogue is only available electronically. The first ten rows of the catalogue are shown here as an example. The columns give the object identification (ID), detector coordinates (X, Y), magnitudes ($V, I, H\alpha$), and corresponding uncertainties ($\sigma_V, \sigma_I, \sigma_H\alpha$).}
	\label{tab:cat}
	\resizebox{\columnwidth}{!}{%
	\begin{tabular}{lcccccccr} 
		\hline
		ID & X & Y & V & $\sigma_V$ & I & $\sigma_I$ & $H\alpha$ & $\sigma _{H\alpha}$ \\
		\hline
   1 & 3720.47 &  949.66 & 20.423 & 0.002 & 19.391 & 0.001 & 19.494 & 0.040\\
   2 & 3729.45 &  960.30 & 22.127 & 0.004 & 21.398 & 0.003 & 21.358 & 0.114\\
   3 & 3599.72 &  982.31 & 26.615 & 0.034 & 25.130 & 0.019 & 22.184 & 0.194\\
   4 & 3676.18 & 1004.65 & 21.246 & 0.003 & 20.213 & 0.002 & 20.343 & 0.061\\
   5 & 3539.25 & 1011.64 & 25.851 & 0.024 & 24.569 & 0.015 & 17.908 & 0.018\\
   6 & 3677.69 & 1026.23 & 22.244 & 0.005 & 21.738 & 0.004 & 21.730 & 0.129\\
   7 & 3677.69 & 1026.23 & 22.244 & 0.005 & 21.738 & 0.004 & 21.730 & 0.129\\
   8 & 3493.73 & 1044.63 & 22.727 & 0.006 & 22.103 & 0.005 & 22.207 & 0.188\\
   9 & 3598.92 & 1049.92 & 21.737 & 0.004 & 20.827 & 0.003 & 20.794 & 0.075\\
  10 & 3547.33 & 1054.39 & 23.654 & 0.009 & 23.052 & 0.007 & 21.964 & 0.149\\
		\hline
	\end{tabular}%
	}
\end{table}
\begin{table}
	\centering
	\caption{Average photometric uncertainties corresponding to different magnitudes in the $V$, $I$, and $H\alpha$ bands. In the $V$ and $I$ bands the uncertainty increases slightly for objects brighter than magnitude 18, due to the short exposures used to study these stars.}
	\label{tab:errors}
	\begin{tabular}{lccr} 
		\hline
		Magnitude & $\sigma_V$ & $\sigma_I$ & $\sigma _{H\alpha}$ \\
		\hline
        12.0 &  -     &  0.001 &  -    \\
        12.5 &  -     &  0.001 &  0.004\\
        13.0 &  0.001 &  0.000 &  0.004\\
        13.5 &  0.002 &  0.002 &  0.003\\
        14.0 &  0.003 &  0.003 &  0.004\\
        14.5 &  0.003 &  0.004 &  0.005\\
        15.0 &  0.004 &  0.005 &  0.005\\
        15.5 &  0.005 &  0.006 &  0.006\\
        16.0 &  0.006 &  0.007 &  0.008\\
        16.5 &  0.008 &  0.009 &  0.010\\
        17.0 &  0.010 &  0.012 &  0.012\\
        17.5 &  0.013 &  0.015 &  0.015\\
        18.0 &  0.001 &  0.001 &  0.019\\
        18.5 &  0.001 &  0.001 &  0.024\\
        19.0 &  0.001 &  0.001 &  0.031\\
        19.5 &  0.001 &  0.001 &  0.040\\
        20.0 &  0.002 &  0.002 &  0.052\\
        20.5 &  0.002 &  0.002 &  0.068\\
        21.0 &  0.003 &  0.003 &  0.092\\
        21.5 &  0.003 &  0.004 &  0.128\\
        22.0 &  0.004 &  0.004 &  0.182\\
        22.5 &  0.005 &  0.006 &  0.264\\
        23.0 &  0.006 &  0.007 &  0.379\\
        23.5 &  0.008 &  0.009 &  0.567\\
        24.0 &  0.010 &  0.011 &  0.865\\
        24.5 &  0.013 &  0.014 &  -\\
        25.0 &  0.016 &  0.018 &  -\\
        25.5 &  0.020 &  0.021 &  -\\
        26.0 &  0.026 &  0.025 &  -\\
        26.5 &  0.032 &  -      & -\\
        27.0 &  0.038 &  -      & -\\
        27.5 &  0.047 &  -      & -\\
        28.0 &  0.058 &  -      & -\\
		\hline
	\end{tabular}
\end{table}
A total of about 58\,700 sources were detected, of which 6\,008 objects exhibit a photometric uncertainty of less than $0.2$ mag in the $H\alpha$ band. As seen in Table~\ref{tab:errors}, the uncertainties are much more significant in $H\alpha$ than in the $V$ and $I$ bands. Therefore, constraining the $H\alpha$ uncertainty also enforces strong error constraints on the $V$ and $I$ bands. In the rest of the paper we consider only these 6\,008 objects.

A visual inspection of the images reveals a large number of resolved galaxies in the background, most of which appear to be well defined. This observation already reveals the low gas density of the field, as well as the low level of extinction, as pointed out by \citet{Sabbi}. These galaxies are the subject of a forthcoming research note (Tsilia et al. in preparation).
\section{Colour--magnitude diagram}

\label{sec:cmd}
\begin{figure}
    \centering\includegraphics[width=0.4\textwidth]{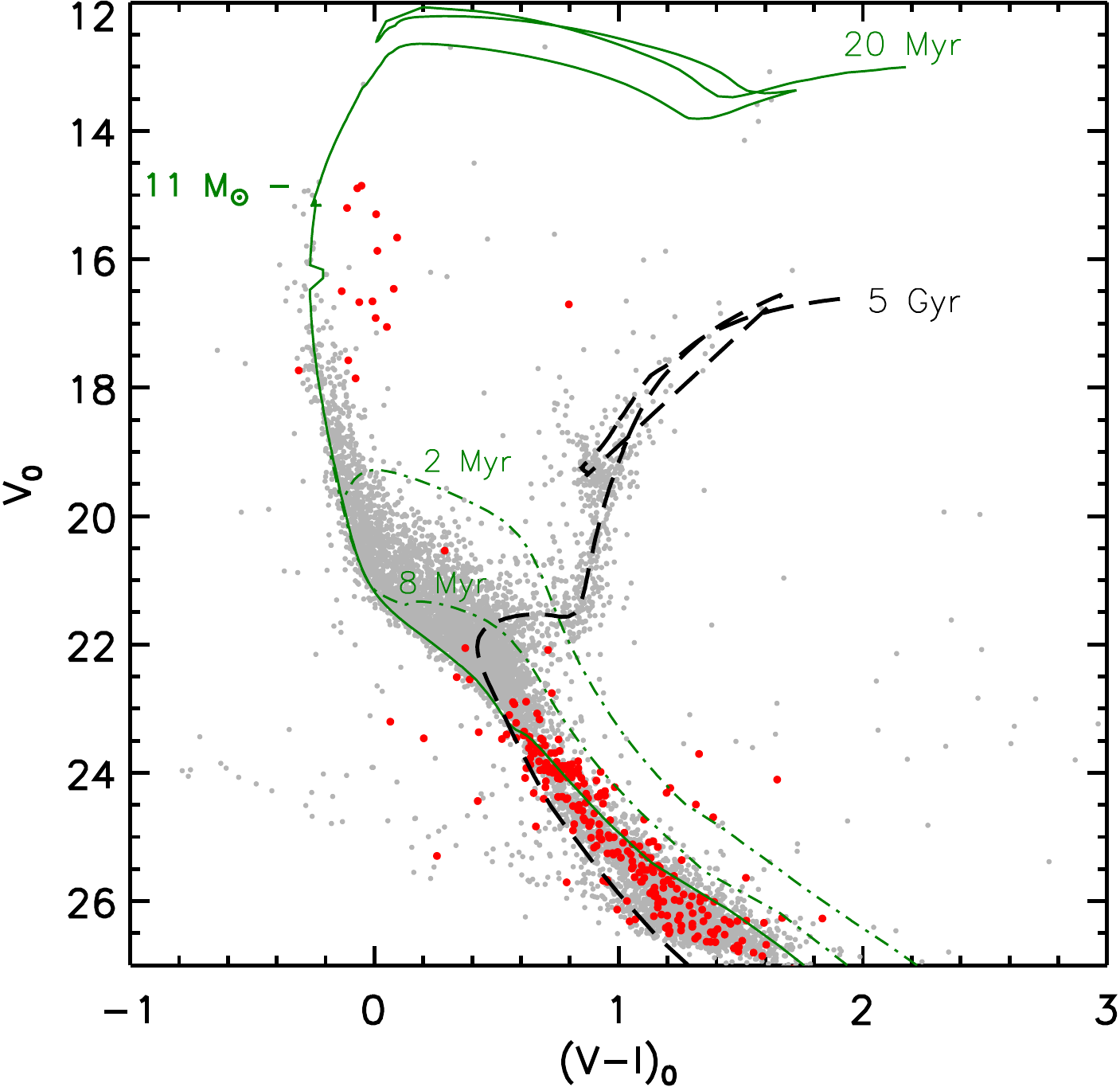}
    \caption{CMD containing all 6\,008 stars with photometric uncertainty of less than 0.2 mag in H$\alpha$. Red dots are stars with H$\alpha$ excess emission above a certain threshold (PMS candidates). Theoretical isochrones are also shown for ages of 2, 8, and 20\,Myr (green) as well as 5\,Gyr (black).}
    \label{fig:cmd}
\end{figure}
Figure~\ref{fig:cmd} shows the CMD in the $V$ and $I$ bands, which was constructed using the 6\,008 objects of appropriate quality. We implemented an extinction correction in order to take into account the foreground extinction along the line of sight caused by the Milky Way. We take the value $E(B-V)=0.07$ as an average of the foreground extinction values measured towards SMC clusters between $0.05$ and $0.09$ (\citealt{piatti2007young}, \citealt{Noel_2007}, \citealt{Schmalzl_2008}). This value is in full agreement with that adopted by \citet{Sabbi}. The standard Galactic extinction law \citep[see][]{1979ARA&A..17...73S,cardelli1989relationship} subsequently gives $A_V=0.18$, $A_I=0.10$, and $A_{H\alpha}=0.12$ for the different filters. The dereddened magnitudes are used in the rest of our analysis.

Regarding the structure of the CMD, we observed both old and young star populations, a prominent main sequence (MS), and many massive stars. The majority of stars in the field of view are old, and many of them are found in the red giant branch \citep[for comparison, we provide a 5\,Gyr isochrone  from the models of][]{tang2014new}. We note that the majority of stars in this diagram are actually old field stars \citep{Sabbi}, and the well-defined shape of the sequences reveals that there is no extinction spread in this field. We also refer the reader to Figure 8\,(c) in \citet{Sabbi}, which contains a CMD of the central NGC\,376 cluster after statistical subtraction of the contamination due to field stars. That figure shows a thin and well-defined MS, revealing that if patchy extinction is present in this field, its effects are negligible.

Young stellar objects constitute the main topic of this work; the most massive ones are located in the upper MS, and include several supergiants. We have drawn the isochrones for 20, 8 and 2\,Myr to guide the eye in Figure~\ref{fig:cmd}, as this is indicative of the ages characterizing these stars; the 2 and 8 Myr isochrones are only shown for the PMS phase and terminate on the MS. \citet{Sabbi} also studied this young population, deriving an age of $28\pm7$\,Myr from isochrone fitting to the massive stars. The location in the CMD of the most massive object still in the MS phase (11\,M$_{\odot}$) is indicated as a reference in the figure. According to \citet{Sabbi}, approximately 90\,\% of the initial mass of the cluster has already been lost, possibly due to gas dispersal following star formation processes. This is possibly another indication of the low density in the environment hosting NGC\,376. The stars with H$\alpha$ excess emission (PMS and massive objects) are addressed in Section~\ref{sec:search}.


\section{Searching for PMS stars}

\label{sec:search}
\begin{figure}
    \centering\includegraphics[width=0.38\textwidth]{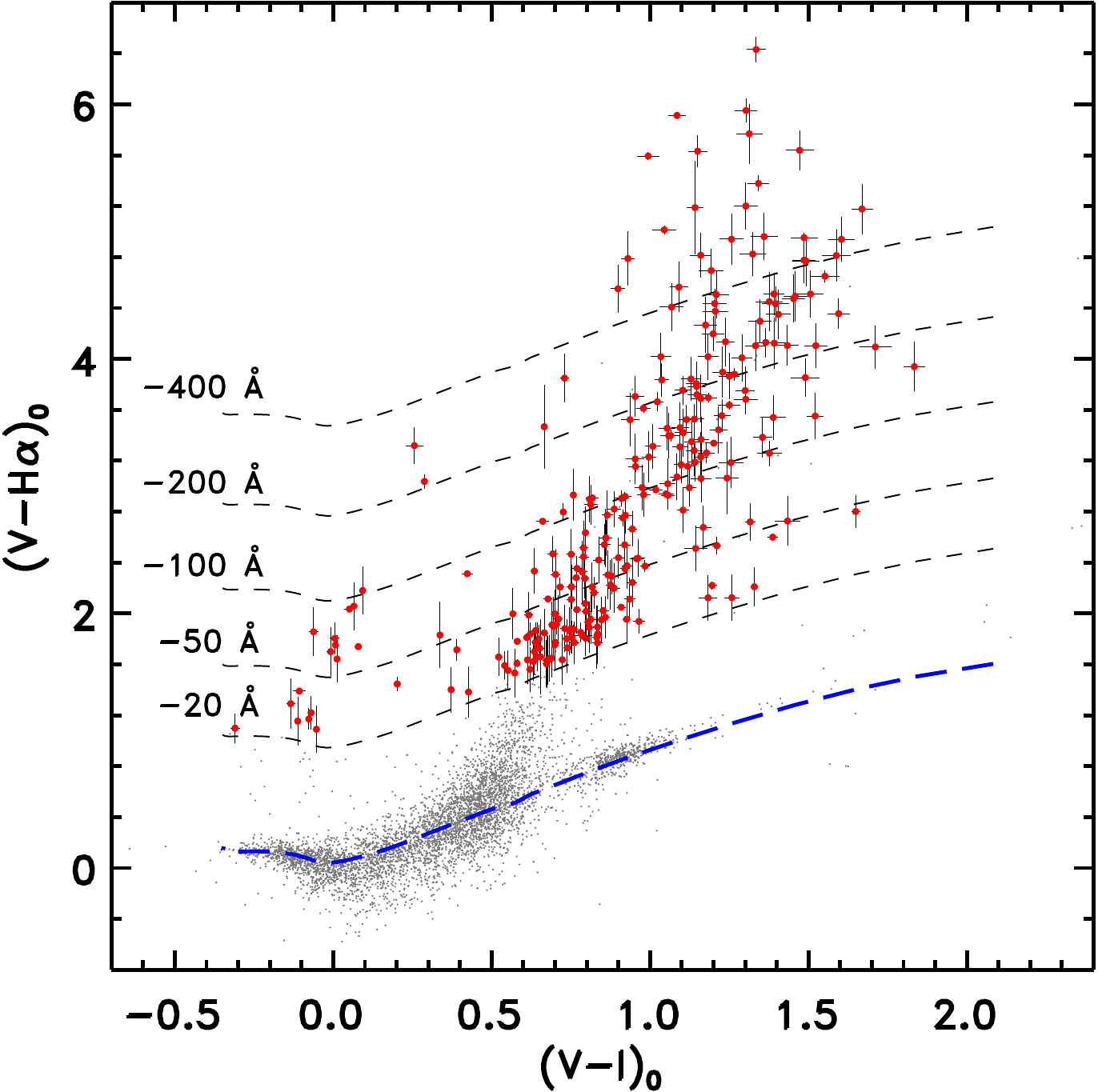}
    \caption{ Identification of objects with excess H$\alpha$ emission. The grey dots refer to the 6\,008 stars with good photometry in all bands. The blue dashed line is the photospheric $V-H\alpha$ colour for the model atmospheres of \citet{bessell1998model} for the metallicity appropriate to the SMC, along which normal stars are expected to be found. The dashed lines of constant equivalent width corresponding to $-20$\,\AA, $-50$\,\AA, $-100$\,\AA, $-200$\,\AA,\ and $-400$\,\AA\ are also shown. Red dots denote stars with H$\alpha$ excess emission higher than the $5\,\sigma$ level and $W_{\rm eq} < -20$\,\AA. The error bars show the uncertainties in the two colours, which are sometimes smaller than the size of the symbols.}
    \label{fig:cc}
\end{figure}

In order to identify candidate PMS objects, we searched for stars exhibiting H$\alpha$ excess emission, which is a characteristic feature of PMS stars \citep[see e.g. the review by][]{calvet}. We built the colour--colour diagram shown in Figure~\ref{fig:cc}, following \citet{De_Marchi_2010,De_Marchi_2011}, and considered a PMS star to be any object that exceeded by $5\,\sigma$ or more the $V-H\alpha$ colour of normal stars of the same effective temperature, as indicated by the $V-I$ colour. 

As \citet{Marchi_2017} showed, the location of ``normal'' non-PMS stars is well represented by the model atmospheres of \citet{bessell1998model} for the specific metallicity of the SMC. Hence, these models were used to identify stars with H$\alpha$ excess emission by comparing the $V-H\alpha$ colour to the colours of normal stars. 

In Figure~\ref{fig:cc} the majority of the stars are normal field objects, and they do not have remarkable H$\alpha$ excess emission.  In order for stars to be classified as candidate PMS objects, we require three conditions to be met. First, the H$\alpha$ excess must be at least five times larger than the combined colour uncertainty of that star. We define the combined colour uncertainty in $V-H\alpha$ as the quantity $\delta_{V-H\alpha}=\sqrt{\delta_V^2+\delta_{H\alpha}^2}$, where $\delta_V$ and $\delta_{H\alpha}$ are the photometric uncertainties in the corresponding bands.
The second condition is that the absolute value of the equivalent width $W_{\rm eq}$ of the H$\alpha$ emission line must be at least 20\,\AA, following the indications of \citet{white2003very}. In this way we exclude older objects with chromospheric activity that could contaminate our sample. Here we adopt the convention that a negative value of the equivalent width corresponds to lines in emission. In order to obtain the equivalent width value for each star, we used the equations from \citet{De_Marchi_2010,De_Marchi_2011}:
\begin{gather}
    W_{\rm eq}=17.679\times(1-10^{0.4\Delta H\alpha})\label{eq1}\\
    \Delta Ha=(V-H\alpha)^{\rm obs}-(V-H\alpha)^{\rm ref}\label{eq2}
\end{gather}
where the superscipt ``obs'' refers to the observed $V-H\alpha$ colour and superscript ``ref'' to the reference template, in our case the $V-H\alpha$ colour that the stars would have if they were found along the theoretical curve. By applying both of these conditions, we identified 264 objects, which are shown as red dots in Figure~\ref{fig:cc}. We ignored stars falling outside the $V-I$ range of the theoretical models, namely with $V-I\leq-0.1$ and $V-I\geq2$.

For the photometric determination of the EW values of our objects with H$\alpha$ excess emission (see Figure~\ref{fig:cc}), this technique was independently verified spectroscopically by \citet{Barentsen2011}. They showed that there are no systematic differences between the EW determined spectroscopically and photometrically for stars in the colour range probed by our observations (see e.g. Figure 5 in that paper). 

The third and final condition to be met in order for stars to be considered candidate PMS objects is that their effective temperature should not exceed 8000\,K. This limit is set because objects of higher temperature are also more massive, while PMS stars are by definition low-mass objects. This temperature corresponds to a colour of $V-I\approx0.5$ in the CMD. We use the candidate PMS objects to explore the physical properties of the young low-mass star population in NGC\,376, which means that photometric completeness is not essential in this project. There are certainly many more candidate PMS stars, starting from the objects in this sample that have smaller amounts of H$\alpha$ excess emission. Furthermore, PMS objects usually exhibit some variability in $H\alpha$ \citep[e.g.][]{joy1945t,appenzeller1989t}, and hence there might be many more PMS stars in the field of view that would have been impossible for us to detect since the data we used cover only one epoch of observation. Therefore, a complete sample was neither needed nor attainable.

\subsection{Spatial distribution}
\label{sec:spatial}
\begin{figure}
    \centering\includegraphics[width=0.4\textwidth]{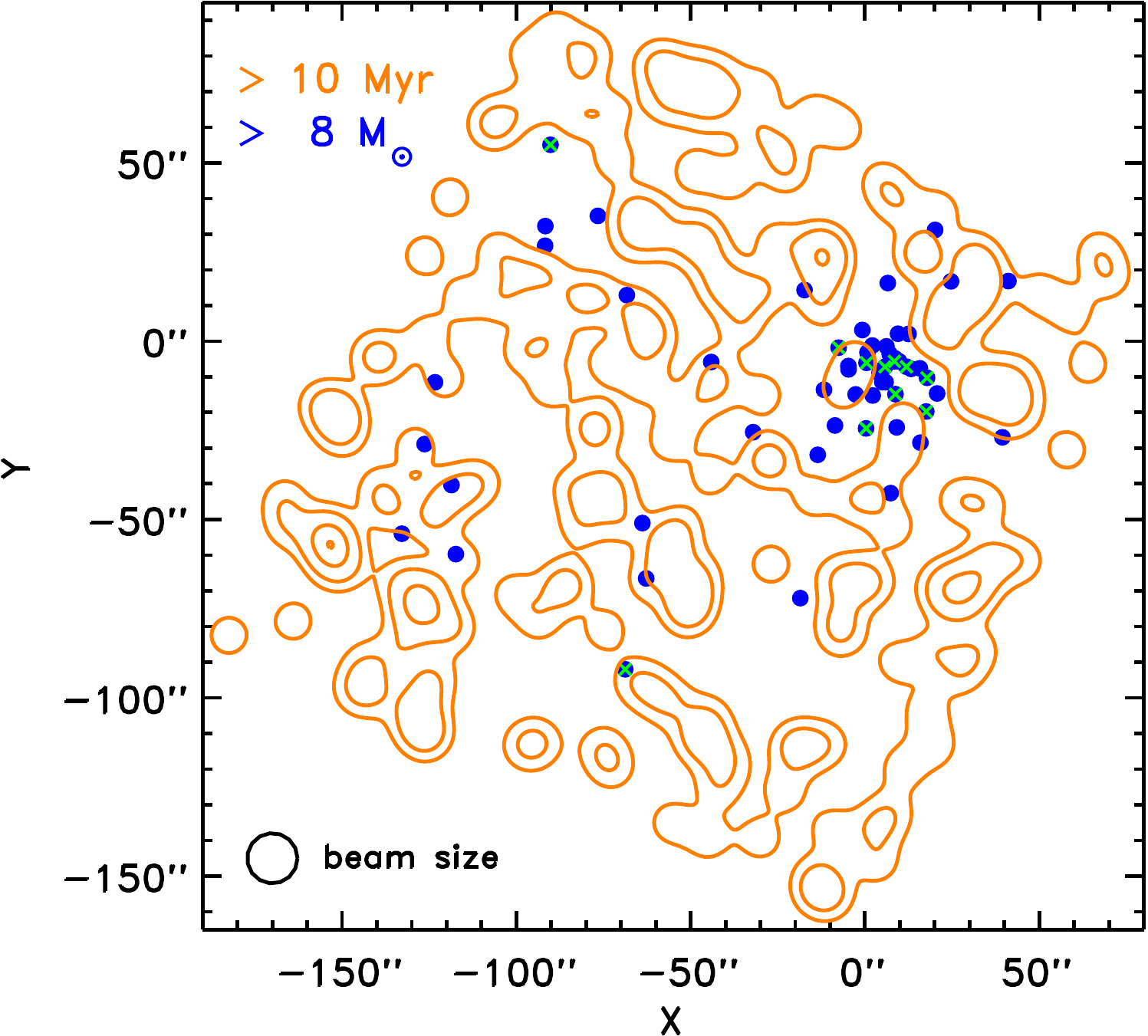}
    \caption{Relative locations of PMS stars (orange contours) and massive MS stars (blue dots). The contour plots were derived after smoothing the distribution with a Gaussian beam whose size is indicated by the circle in the lower left corner ($\sigma = 4\arcsec$). The step between contour levels corresponds to a factor of $1.5$. For reference, the blue dots denote the massive upper MS stars in the field ($> 8$\,M$_{\odot}$). The massive stars exhibiting H$\alpha$ excess emission are indicated by an additional cross. As discussed in Section~\ref{sec:dist}, these are possibly Be stars. Unlike massive stars, PMS objects are rather uniformly distributed across the field and do not reveal any central concentration.}
    \label{fig:spatial}
\end{figure}

The contour plots in Figure~\ref{fig:spatial} show the distribution of PMS stars in the field of view. The figure shows that low-mass PMS stars are distributed rather evenly across the field, while the massive stars are found to be more concentrated around the centre of the cluster.   

Some small concentrations of high-mass stars are also observed further away from the cluster centre. Stars in these areas might have been ejected from NGC\,376, as already discussed by \citet{Sabbi}, but they could also suggest the existence of other smaller unrelated star-forming areas around the main NGC\,376 cluster. Detailed information on the velocity of massive stars, obtained through spectroscopy, would be needed to investigate more in depth the kinematics and likely dynamical history of these objects.
If the low-mass PMS stars formed all in NGC\,376, their uniform distribution suggests that they might have been ejected or have migrated out of the cluster through gravitational interactions. Considering the size of the field of view of about 48\,pc (see Figure 1), a velocity dispersion of only 2\,km\,s$^{-1}$ would allow PMS stars to traverse about 56\,pc in 28\,Myr, which is the age of the cluster calculated by \citet{Sabbi}. This would naturally account for the observed spatial distribution, so it is possible that all PMS stars formed near the centre of the cluster and then migrated outwards. This is also in agreement with the cluster's escape velocity, estimated by \citet{Sabbi} to be only 3\,km\,s$^{-1}$.

A markedly different spatial distribution between younger and older stars with H$\alpha$ excess has been consistently observed in all the regions studied in this way. This includes: stars in the field of SN\,87A \citep{De_Marchi_2010}, in NGC\,346 \citep{De_Marchi_2011}, in 30 Dor \citep{DeMarchi2011, Marchi_2017} and its surroundings \citep{10.1111/j.1365-2966.2011.20130.x}, in NGC\,602 \citep{DeMarchi2013}, in NGC\,3603 \citep{2010AAS...21546329B}, in Trumpler 14 \citep{Beccari2015}, in LH\,95 \citep{biazzo2019photometric}, as well as stars in the NGC\,6611 cluster (Eagle nebula) studied on the basis of their X-ray luminosities, near-infrared excess, and Li absorption features \citep{De_Marchi_2013}. In all these cases the younger stars are more centrally concentrated, while the older stars are more uniformly distributed across the field. Three of these clusters (NGC\,3603, NGC\,6611, Tr\,14) are in the Milky Way. 

\subsection{Distribution in the colour--magnitude diagram}
\label{sec:dist}
The position of the PMS stars in the CMD indicates that they constitute a relatively old population. In Figure~\ref{fig:cmd} objects with H$\alpha$ excess emission (red dots) are typically close to the MS, which indicates that these are older PMS stars. Younger PMS objects would be considerably redder and brighter than MS stars in these colours, and very few of them are seen in the CMD. We note that old populations of PMS stars are not unusual \citep[see e.g.][]{De_Marchi_2011,Marchi_2017}. 

To guide the eye, we also include in Figure~\ref{fig:cmd} the PMS isochrones for ages of 2 and 8\,Myr (dot-dashed lines) and\,20 Myr (solid line) from the models of \citet{chen2015parsec}, for metallicity $Z=0.004$, in agreement with the currently accepted metallicity values for the SMC \citep[see][]{russell1992abundances,rolleston1999chemical,lee2005chemical,perez2005comparative}. In Section~\ref{sec:phys} we derive  the stellar parameters of these candidate PMS objects, including their ages, but the comparison with the isochrones already suggests an age greater than 20\,Myr for most of these PMS stars.

In principle, if the reddening correction were excessive it could artificially increase the derived age of the PMS stars. However, we applied the same small correction to all objects in the field, namely $E(V-I)=0.08$, and even if no reddening correction had been applied most of the PMS objects would still be bluer than the 20\,Myr isochrone.

Although a circumstellar disc seen at high inclination could make the colour of PMS stars appear bluer than they intrinsically are, this would statistically only affect at most a small percentage of the PMS stars \citep{De_Marchi_2013}. The conclusion, therefore, is that the vast majority of stars with H$\alpha$ excess emission in this field are older than $\sim 20$ Myr. 

We note that with broad-band photometry alone it would not have been possible to identify these PMS objects because their colours would not distinguish the older PMS stars from MS objects. It is only the combination of broad-band and narrow-band photometry that allows us to identify this old population by searching for objects with H$\alpha$ excess emission.

In Figure~\ref{fig:cmd} we also observe some massive stars with H$\alpha$ excess emission, located near the upper MS. These are quite possibly Be stars \citep[e.g.][]{Rivinius_2013}. These stars are characterized by excess emission in the Balmer series (particularly in H$\alpha$, which is what we see), variability, and typical values for the emission equivalent width in H$\alpha$ of some tens of \AA, which are perfectly in line with those that we find. For instance, \citet{refId0} use slitless spectroscopy to study Be stars in the SMC cluster NGC\,330 and measure $H\alpha$ EW values in the range $\sim 20-80$\,\AA, in excellent agreement with our values in Figure\,\ref{fig:cc}. This gives us further confidence in the accuracy of our photometric determination of the EW. Our massive stars with excess emission typically exhibit $0 < V-I < 0.2$ before the reddening correction is applied, a colour range again fully in agreement with the expected $V-I$ colours of Be stars in the SMC \citep{10.1093/mnras/stab1481}. 

Alternatively, these objects could also be Herbig Ae/Be stars \citep[namely massive PMS stars, see][]{1997SSRv...82..407P,waters1998herbig}, but as we show later, in light of the mass of the stars, the implied very young age of such objects would be at odds with the general age of the cluster. If these stars were Herbig Ae/Be objects, they would be younger than $\sim 250\,000$\,yr, but no other such young objects are observed in the field. So this interpretation, although plausible, appears unlikely. Thus, the most likely conclusion is that the massive stars with excess are indeed Be stars. To validate this conclusion one would need spectroscopic observations, in order to properly characterize the spectral features of these stars, as well as some measurement of their photometric or spectral variability. 

We exclude these massive stars exhibiting H$\alpha$ excess emission from the rest of our analysis, so the sample of our candidate objects is reduced to 244 stars. In the next section we derive their physical properties.

Contamination in this sample of PMS candidate objects is negligible. The frequency of other types of stars with H$\alpha$ excess emission close to the MS, where most of our candidates are located, is extremely low. Their typical fraction is found to be approximately 0.001\%, according to the IPHAS survey \citep{10.1111/j.1365-2966.2005.09330.x, 10.1111/j.1365-2966.2007.12774.x}. Only regions of current or recent star formation were found to exhibit a higher density of observed H$\alpha$ emitters.


\section{Physical properties of candidate PMS stars}
\label{sec:phys}
In order to derive the physical properties of our sample of 244 candidate PMS stars, we used dereddened colours and magnitudes to obtain information on their mass, age, luminosity, and mass accretion rate. To derive individual masses and ages, we compared the position of each star in the Hertzsprung--Russell diagram (HR) with PMS evolutionary tracks, in order to determine the most likely age and mass for each object. The first step is the determination of the effective temperature $T_{\rm eff}$ and bolometric luminosity $L$ from the photometry. The value of $T_{\rm eff}$ was derived from the dereddened $V-I$ colour, which is a very reliable index for temperature determinations in the range 4000--8000\,K of interest here \citep[e.g.][]{bessell1998model,phd}. The models of \citet{bessell1998model}, already discussed in Section~\ref{sec:search}, were used in the conversion. The value of $L$ was obtained from the dereddend $V$ magnitude and the same \citet{bessell1998model} models, with an adopted distance modulus of $18.91$. The uncertainties on $T_{\rm eff}$ and $L_{\rm bol}$ are dictated by the accuracy of the photometry. The typical uncertainty on $T_{\rm eff}$ is about 50\,K and does not vary significantly in the range $4000 - 6000$\,K, where most of our sources fall, because of the rather constant uncertainties on the $V-I$ colour. On the other hand, the uncertainty on $L_{\rm bol}$ depends on the luminosity itself, but the correlation is rather tight and the relative uncertainty on $L_{\rm bol}$ amounts to $11\,\%$.

The comparison between observations and models is accomplished as follows. A grid is defined in luminosity and temperature in the HR diagram, using logarithmic steps and evenly spaced cells whose size is comparable to the observational uncertainties of each star. Then the evolutionary models of \citet{tognelli2011pisa} for metallicity $Z=0.004$ are used, interpolated to a fine mass grid with a logarithmic step of $0.025$. We determine all of the cells in our HR grid that a star of a given mass should cross throughout its evolution, according to the corresponding evolutionary track for that mass. We calculate, for all of the grid cells, the crossing time (difference between exit and entrance times) and characteristic age (average of exit and entrance times), and complete this process for all evolutionary tracks.

As a result, we obtain four matrices containing information regarding the tracks that cross each cell, more specifically the mass, age, time of permanence in the cell, and multiplicity for that cell (number of possible solutions). In the case of multiple solutions, weights are assigned to each one. The most likely mass and age of each individual observed star is given by the cell of the grid to which the star belongs.

This method was originally presented by \citet{phd} and then also successfully used in more recent works on PMS stars \citep[see e.g.][]{De_Marchi_2011,De_Marchi_2013,Marchi_2017}. It does not require any further assumptions on the properties of the stellar population as a whole.

As pointed out by \citet{Marchi_2017}, in typical star-forming regions the main source of uncertainty on the ages and masses derived in this way is the reddening correction since interstellar extinction typically shifts stars in the HRD towards lower masses and younger ages. However, in the case of NGC\,376 there is no detectable differential extinction in this field (see Section 3) and the only correction applied is to remove the foreground Galactic reddening $E(B-V)=0.07$. The typical dispersion around this value, as mentioned above, is about $0.02$\,mag. Combined with the photometric uncertainties of the PMS star candidates, this implies an overall uncertainty not exceeding 4\,\% on the derived masses of 8\,\% on the relative ages. Absolute values are of  course affected by any uncertainties inherent in the theoretical evolutionary tracks used for the comparison in the HRD, but, as already pointed out, this does not hinder the relative comparison of our PMS candidates.

\begin{figure}
    \centering
    \begin{subfigure}[t]{0.245\textwidth}
        \centering
        \includegraphics[width=\linewidth]{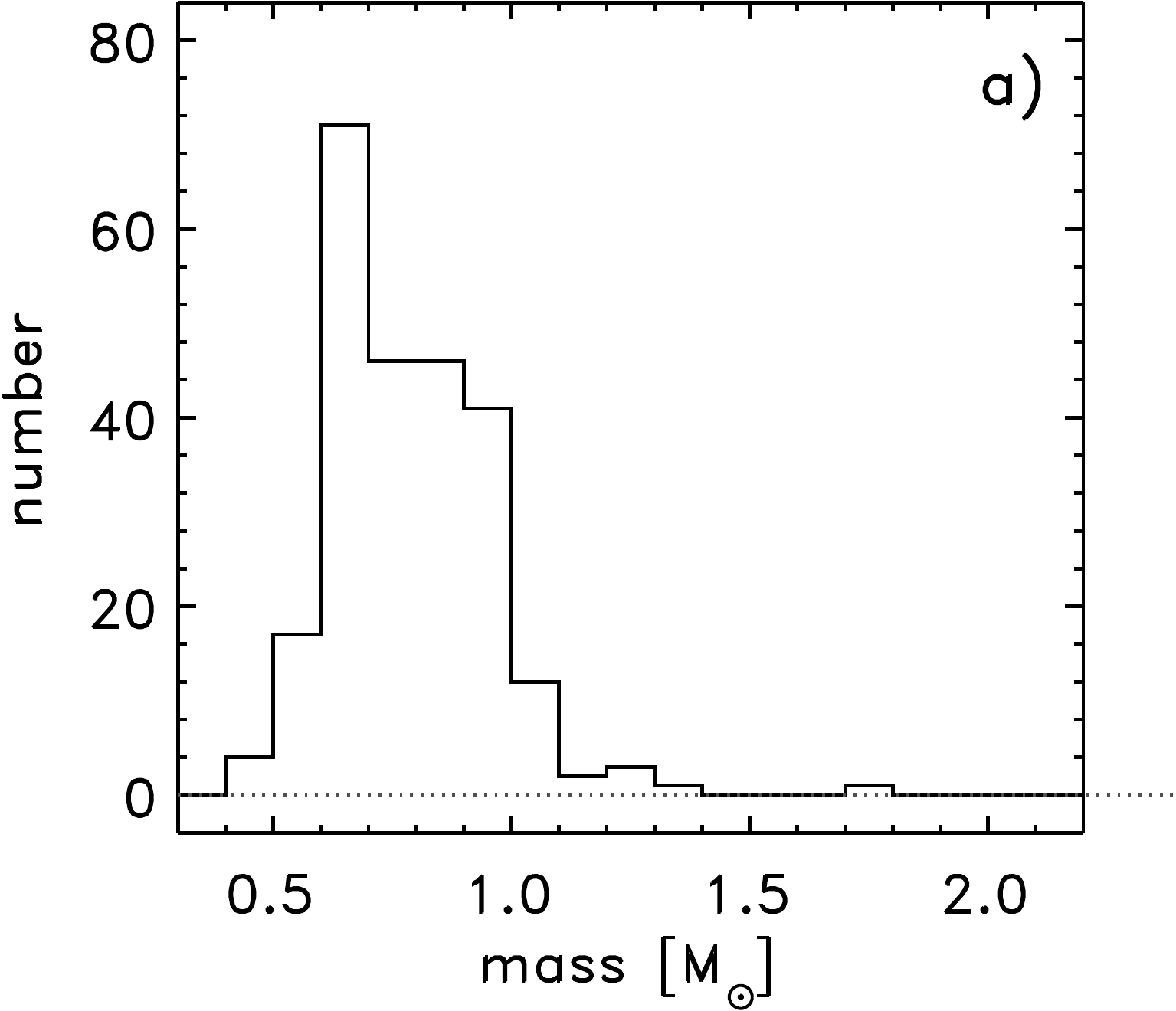} 
        \phantomcaption
        \label{fig:histmass}
    \end{subfigure}
    \hfill
    \begin{subfigure}[t]{0.23\textwidth}
        \centering
        \includegraphics[width=\linewidth]{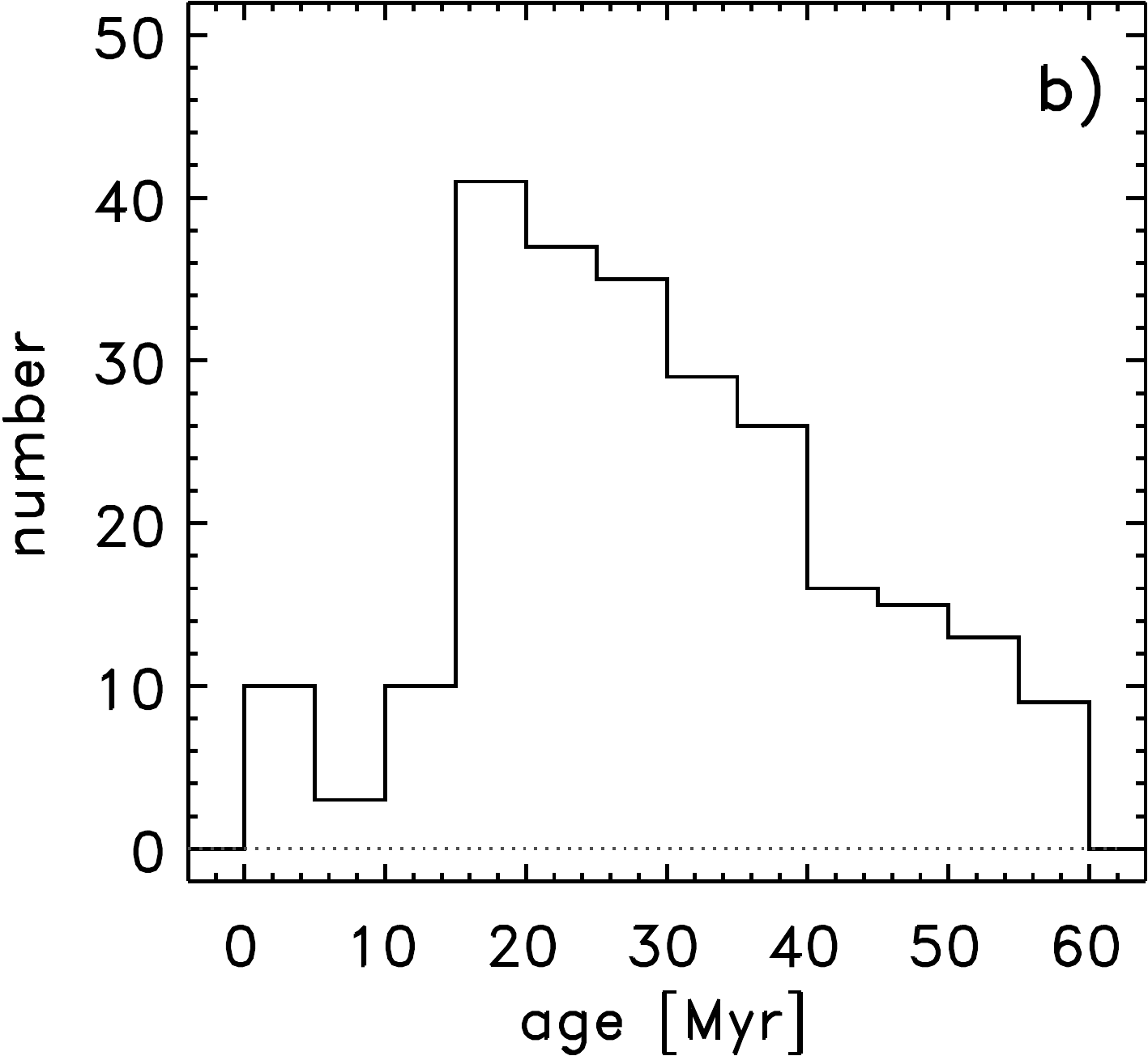} 
        \phantomcaption
        \label{fig:histage}
    \end{subfigure}
    \begin{subfigure}[t]{0.235\textwidth}
        \centering
        \includegraphics[width=\linewidth]{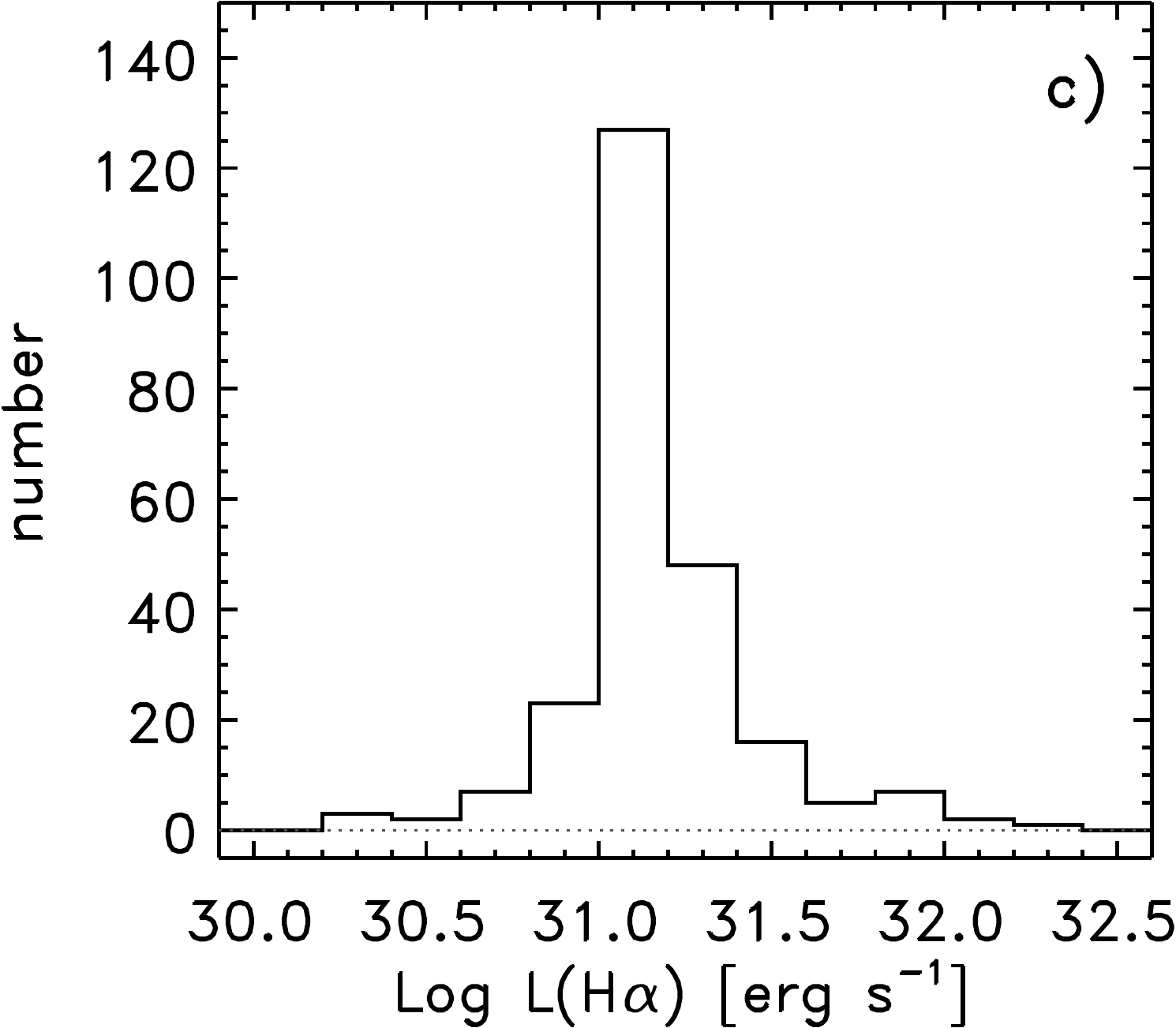} 
        \phantomcaption
        \label{fig:histlum}
    \end{subfigure}
        \hfill
    \begin{subfigure}[t]{0.23\textwidth}
        \centering
        \includegraphics[width=\linewidth]{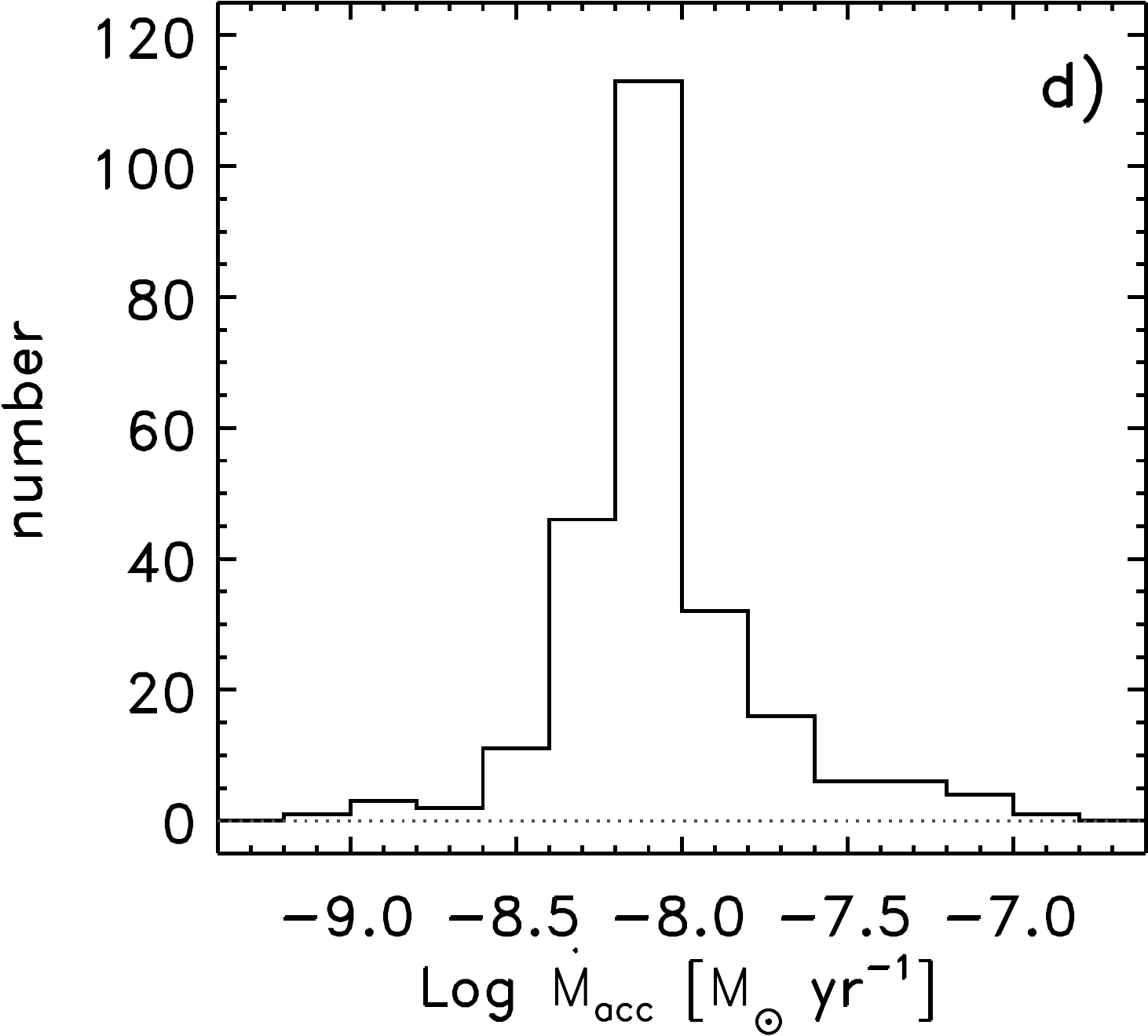} 
        \phantomcaption
        \label{fig:histmdot}
    \end{subfigure}
    \caption{All panels refer to our sample of 244 candidate PMS stars and show the distribution of masses (Panel a), ages (Panel b), $L(H\alpha)$ values (Panel c), and mass accretion rate values (Panel c).}
\end{figure}

We then constructed histograms of mass and age, shown in \Cref{fig:histmass} and \Cref{fig:histage}, respectively. We find a mean mass of $0.79\pm0.17$\,M$_{\odot}$, which is within the range we expected for the low-mass stars that were the target of our study. Since the objects showing signs of accretion in this field are almost all less massive than 1\,M$_{\odot}$, and accretion from a circumstellar disc in the PMS phase is a sign of youth (massive PMS stars terminate the accretion phase more quickly and reach the MS in less time compared to lower-mass stars), we conclude that this stellar population is indeed at an advanced PMS stage, with low-mass stars still reaching the MS but more massive objects having already entered the MS phase. As already mentioned, had it not been for the H$\alpha$ excess emission, we would not have been able to distinguish these older PMS stars from the much more abundant MS stars. Although the relative mass distribution of PMS stars is shown, it is important to understand that this cannot constrain the shape of the mass function since we are only looking at stars that at the time of the observations had H$\alpha$ excess emission and we are not considering the unavoidable photometric incompleteness.

The median value for the age is found to be 28\,Myr, with ages covering a broad range from about 20 to 40\,Myr (25th and 75th percentiles, respectively). The median age effectively confirms the results of \citet{Sabbi}, who studied massive objects in the same field. This is an important result since our work, an independent study of a different type of objects (i.e. low-mass stars), reached the same conclusion. Interestingly, we do not find signs of a younger generation of PMS stars, which appears somewhat unusual in the Magellanic Clouds. Regions previously studied consistently revealed multiple generations \citep[see e.g.][]{De_Marchi_2011,De_Marchi_2013,Marchi_2017}. One possible interpretation is that an insufficient amount of gas was available to form new stars. As \citet{Sabbi} showed, as much as 90\,\% of the original mass of the cluster appears to have already been expelled, and NGC\,376 is, in fact, slowly dissolving into the field of the SMC.

Next we derive the accretion luminosity from the H$\alpha$ luminosity. The H$\alpha$ emission line luminosity $L(H\alpha)$ for the PMS stars is found from the $\Delta H\alpha$ colour (see equation~\ref{eq2}), which corresponds to the net excess emission in the H$\alpha$ line. $\Delta H\alpha$ is defined as the difference between the $V-H\alpha$ colour of PMS and normal objects with the same $T_{\rm eff}$, as discussed in Section~\ref{sec:search}. The value of $L(H\alpha)$ is then obtained taking into account the distance modulus for NGC\,376 (which we took from \citealt{Sabbi} to be $18.91 \pm 0.04$) and the sensitivity of the F656N filter \citep[for details, see][]{De_Marchi_2010,De_Marchi_2011,De_Marchi_2013}.
The corresponding histogram is shown in Figure~\ref{fig:histlum}. The $L(H\alpha)$ values are characterized by a median of $10^{-31.1}$\,erg s$^{-1}$ (i.e. $0.0036$ L$_{\odot}$). We note that, compared with the case of the SMC cluster NGC\,346 \citep[see][]{De_Marchi_2011}, the peak in the distribution of $L(H\alpha)$ for NGC\,376 is about a factor of $2.5$ lower. The difference, however, is understandable, since NGC\,346 is presently the region of most intense star formation in the entire SMC. It contains over 30 O-type stars \citep{masseyetal1989,evansetal2006} with estimated ages of $\sim 1-3$\,Myr) and a rich population of equally young PMS stars \citep{notaetal2006,sabbietal2007,gouliermisetal2007,hennekemperetal2008,De_Marchi_2011}. On the other hand, since in NGC\,376 we have on average older and less massive PMS stars, their typical accretion luminosity, and hence $L(H\alpha)$, is expected to be weaker.

The accretion luminosity measures the energy released in the last step of the recombination process of hydrogen atoms that had been ionized by gravitational energy released during the accretion process \citep[e.g.][]{1998ApJ...495..385H}. In order to derive the accretion luminosity we follow \citet{De_Marchi_2010,Marchi_2017}, who give $\log L_{acc} =(1.72\pm0.25) + \log L(H\alpha)$, where the uncertainty is the same for all objects. This relationship was based on \citet{Dahm2008} and a small dataset of measurements, which is now superseded by the richer sample of simultaneous measurements of multiple emission lines for 45 young stellar objects in Lupus \citep{Alcal2017}. Using this more recent dataset, we derived a more solid linear relationship that best fits the data, which is slightly less steep, $\log L_{acc} = (1.38\pm0.25) + \log L(H\alpha)$, and will be discussed in a future paper (De Marchi et al. in preparation). Within the uncertainties the two relationships are in agreement. Due to the systematic nature of this effect, it is important to take it into account, but at the same time the new relationship will not affect the differences in the accretion properties of stars in different regions. Therefore, for ease of comparison with previous works, in this paper we continue to use the original relationship.

From $L_{\rm acc}$ we derive the mass accretion rate. Its calculation requires a measure of the stellar photospheric radius, which comes from the absolute magnitude
\begin{equation}
R_*=\sqrt{10^{-0.4\Delta V}}\times\num{6.96e10},
\end{equation}
where $\Delta V=V-(m - M)_0-V^{ \rm ref}$, the distance modulus to the SMC is $(m - M)_0= 18.91\pm0.04$ (as mentioned above), and $V^{\rm ref}$ is the theoretical absolute $V$ magnitude of the star derived by interpolating along the \citet{bessell1998model} models. The actual value of the mass accretion rate follows from the free-fall equation that links the luminosity released in the impact of the accretion flow with the rate of mass accretion $\dot M_{\rm acc}$ through the relationship
\begin{equation}
L_{\rm acc} \simeq \frac{G\,M_*\,\dot M_{\rm acc}}{R_*} \left(1 -
\frac{R_*}{R_{\rm in}}\right),
\end{equation}
where $G$ is the gravitational constant, $M_*$ is the stellar mass, and we assume the internal radius of the circumstellar disc to be $R_{\rm in} = 5 R_*$, following \citet{gullbring1998disk}. The main source of uncertainty on $\dot M_{\rm acc}$ is the conversion between $L(H\alpha)$ and $L_{\rm acc}$ discussed above. \citet{De_Marchi_2011} showed that the intrinsic spread on this relationship does not exceed $0.25$\,dex, which we adopt as characteristic uncertainties for our $\dot M_{\rm acc}$ values.

The histogram of the mass accretion rate is shown in Figure~\ref{fig:histmdot}. It is interesting to note that while NGC\,346 exhibits a broad peak from $10^{-8}$ to $10^{-7}$ M$_{\odot}$\,yr$^{-1}$ \citep[see][]{De_Marchi_2011}, in NGC\,376 we observe a typical lower mass accretion rate with a median of $7.9 \times 10^{-9}$  M$_{\odot}$\,yr$^{-1}$. We note once again that NGC\,376 contains a much smaller number of massive and young PMS stars than NGC\,346 and this is expected to contribute to this result, although the characteristic mass and age difference with respect to NGC\,346 does not appear to be sufficient to fully account for the stark discrepancy in the values of $\dot M_{\rm acc}$.
\begin{figure}
    \centering\includegraphics[width=0.45\textwidth]{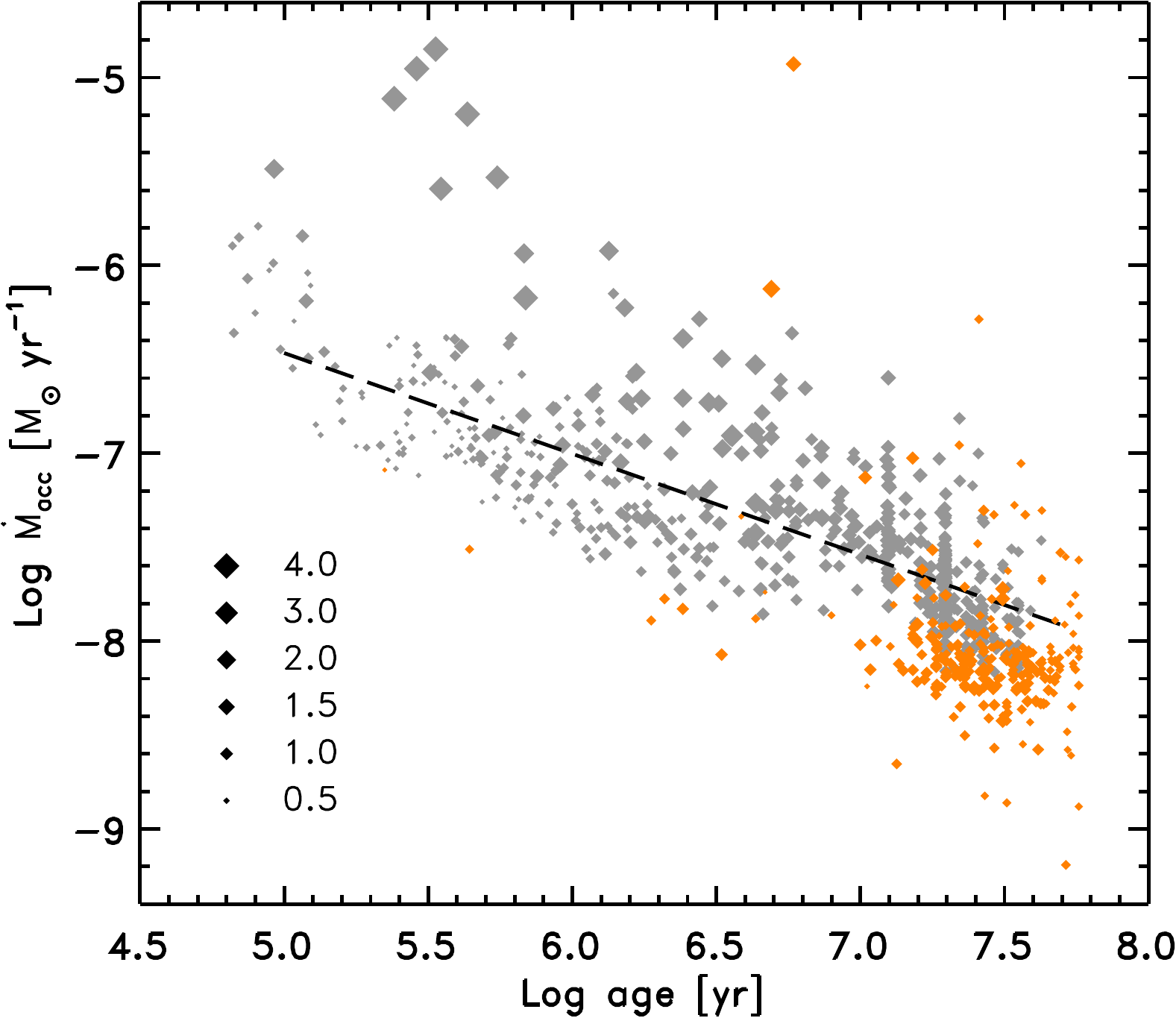}
    \caption{Mass accretion rate of the NGC\,346 (in grey) and the NGC\,376 (in orange) candidate PMS stars as a function of age. The dashed line represents the best fit to the distribution in NGC\,346. The size of the symbols is proportional to the mass of the stars, indicated in the legend in solar units.}
    \label{fig:346}
\end{figure} 
To highlight this, we show in Figure~\ref{fig:346} a more direct comparison of the mass accretion properties in NGC\,376 and NGC\,346, where $\dot M_{\rm acc}$ is plotted as a function of the age of the PMS stars. These clusters are characterized by a similar metallicity \citep[see e.g.][]{russell1992abundances,rolleston1999chemical,lee2005chemical}. However, the stars in NGC\,376 appear to have a systematically lower $\dot M_{\rm acc}$ value, even when the comparison is limited to stars with similar ages and masses. To guide the eye, the dashed line shows the best fit to the distribution observed in NGC\,346. The mass accretion rate of objects older than 10\,Myr, which are common to both regions, is $0.4$\,dex or a factor of $2.5$ lower in NGC\,376. The median value of the mass of the NGC\,376 PMS stars in this age range ($\sim 0.75$\,M$_{\odot}$) is only $\sim 30\,\%$ lower than that for objects with similar ages in NGC\,346 (median mass $\sim 1.05$\,M$_{\odot}$), and therefore cannot account for the difference observed in the mass accretion rate.

\citet{Marchi_2017} showed that the mass and age of a forming star must be considered simultaneously when studying accretion processes because the range of masses of objects detectable as PMS stars shrinks over time and more massive PMS objects that have reached the MS no longer accrete. To determine reliable correlations between these parameters, \citet{Marchi_2017} used a multivariate linear regression fit of the type $\log \dot M_{\rm acc} = a \times \log t + b \times\log m + c$. If the mass accretion rate depended solely on the mass and age, the $c$ term would have to exhibit the same value across the different star-forming regions that they studied. Instead, this term was shown to vary, and these authors discovered a correlation with metallicity $Z$ whereby in lower-metallicity environments PMS stars accrete more and for a longer time \citep[for details, see][]{Marchi_2017}. A possible explanation of this result is that a forming star exerts lower radiation pressure on low-metallicity circumstellar disc material. Therefore, in those environments the disc is dispersed less efficiently, and accretion may remain active for a longer period of time.

The correlation between $c$ and $Z$ that \citet{Marchi_2017} found (see their Figure 11) exhibits some dispersion, which these authors postulate might be related to the density of the region. This is suggested by the systematically lower value of $c$ in one of the clusters that they studied, NGC\,602, a lower-density star-forming region in the Wing of the SMC. The same effect has recently been observed in LH\,95, a low-density region of star formation in the Large Magellanic Cloud \citep{biazzo2019photometric}.

For the sake of comparison, we applied the same multi-parametric fit to the NGC\,376 objects, adopting the same values for the $a$ and $b$ parameters over the same range of mass and ages. We found in this way the value of $c$ for NGC\,376 to be $-3.65\pm0.05$, which is considerably lower than the value of $-3.41 \pm 0.03$ characteristic of NGC\,346 at a similar metallicity.

Metallicity alone cannot explain this result since it is similar for the two clusters; therefore, as \citet{Marchi_2017}, we conclude that another environmental factor appears to have an impact on the mass accretion rate. Again, gas density in NGC\,376  appears to be lower than in NGC\,346, owing to the fact that NGC\,376 is located in the outskirts of the SMC, in what is known as the SMC Wing, well outside the main body of the galaxy, at a projected distance of about 850\,pc from its centre. A clear indicator of the lower density of the NGC\,376 environment is given not only by its projected position on the sky, but even more objectively by the large number of background galaxies; as we pointed out, there are over 90 in this field, suggesting that the region is generally transparent (i.e. it lacks large-scale dust and molecular concentrations). Conversely, an inspection of the area of the same size containing NGC\,346, observed with the same instrumentation (ACS on board the HST) and similar exposure times (proposal ID 10248, principal investigator: A. Nota), only reveals about a dozen background galaxies, since most of the region is affected by extended nebulosity. So it is possible that the lower density of the NGC\,376 environment (as in the cases of NGC\,602 and LH\,95 mentioned above) is responsible for a lower effective mass accretion rate. For instance, although the efficiency of star formation in low-density environments is subject to some debate, there is evidence that it may be reduced compared to that in massive clusters \citep{adamo2015imprints}. While the mechanism for the accretion of matter onto PMS stars in these environments is not known and will require further investigation, it is conceivable that if the circumstellar discs are less rich than in denser environments, the rate of mass accretion will also be reduced.


\section{Summary and conclusions}

\label{sec:sum}
In this work we searched for PMS stars in NGC\,376 in the SMC, and studied their physical properties. We combined broad-band photometry in the $V$ and $I$ bands with narrow-band imaging in $H\alpha$. The observations were obtained with the ACS and WFC3 instruments on board the HST. The main results of our study can be summarized as follows:
\begin{enumerate}
\item We constructed a CMD that reveals both young and old stellar populations, with the majority of them being actually old SMC field stars.
\item We constructed a ($V-H\alpha$, $V-I$) colour--colour diagram and identified 244 candidate PMS stars, which are objects exhibiting emission in the $H\alpha$ band exceeding by at least $5\,\sigma$ that of normal stars with the same effective temperature, with an absolute value of the equivalent width of the H$\alpha$ line larger than 20\,\AA.
\item We also identified a population of massive stars with H$\alpha$ excess emission, which are likely to constitute Be stars.
\item Most of the PMS stars that we find appear to be older than $\sim 20$\,Myr, and in the CMD they are located close to the MS. 
\item An isochrone fit suggests that the median age of the PMS stars in NGC\,376 is 28\,Myr, with the 25th and 75th percentiles of the age distribution respectively at $\sim 20$ and $\sim 40$\,Myr. This is in excellent agreement with the results from \citet{Sabbi} on the age of massive stars in this cluster ($28\pm7$\,Myr). These are independent measurements of two very different kinds of objects, low-mass and high-mass stars, that reach the same conclusion, giving us confidence in the validity of our investigation.
\item We examined the spatial distribution of stars across the field of view, and found that low-mass PMS stars are uniformly distributed, while the most massive stars are more centrally concentrated. Nevertheless, the two distributions are fully consistent with the scenario of central formation and consequent migration when a velocity dispersion of at least 1\,km\,s$^{-1}$ is assumed, considering the age of the cluster. This is fully in line with the cluster's escape velocity estimated by \citet{Sabbi} to be only 3\,km\,s$^{-1}$. 
\item Through comparison with theoretical PMS models we also determined the mass of these PMS stars, finding a mean value of $0.79 \pm 0.17$\,M$_{\odot}$. We also derived the H$\alpha$ luminosity, accretion luminosity, and mass accretion rate of these objects. 
\item We studied the mass accretion rate in NGC\,376 as a function of the mass and age of the PMS stars and compared it with the same parameters measured in NGC\,346, a massive star-forming region in the SMC sharing the same metallicity as NGC\,376. 
\item We find that the mass accretion rate is systematically lower in NGC\,376 than in NGC\,346 even when objects of the same mass and age are compared. We attribute this difference to other environmental factors, the most obvious of which is the lower gas density in NGC\,376 revealed by the number of background galaxies in this field, which is approximately eight times greater than in NGC\,346.
\item In conclusion, our research suggests that, in addition to the mass, age, and metallicity of PMS stars, the prevailing gas density of the star-forming region may also have an effect on the rate of mass accretion, with lower accretion rates in lower-density environments.
\end{enumerate}

Spectra are needed to fully characterise the physical nature of the PMS candidates in NGC\,376, but no instrument to date has offered the necessary combination of sensitivity and spatial resolution required to study PMS stars in another galaxy. With the advent of JWST the situation has now changed, and the Near Infrared Spectrograph onboard JWST will finally allow the complete characterisation of the accretion properties of these PMS objects in a low-metallicity environment. A preliminary study of the spectra of PMS stars in the NGC\,346 cluster in the SMC \citep{demarchi2023} suggests that at low metallicity the lifetime of circumstellar discs is longer than for similar objects in the solar neighbourhood.


%
   \bibliographystyle{aa} 
   \bibliography{references} 

\begin{thebibliography}{63}
\expandafter\ifx\csname natexlab\endcsname\relax\def\natexlab#1{#1}\fi

\bibitem[{Adamo(2015)}]{adamo2015imprints}
Adamo, A. 2015, Proceedings of the International Astronomical Union, 12, 17

\bibitem[{Alcal{\'{a}} {et~al.}(2017)Alcal{\'{a}}, Manara, Natta, Frasca,
  Testi, Nisini, Stelzer, Williams, Antoniucci, Biazzo, Covino, Esposito,
  Getman, \& Rigliaco}]{Alcal2017}
Alcal{\'{a}}, J.~M., Manara, C.~F., Natta, A., {et~al.} 2017, Astronomy {\&}
  Astrophysics, 600, A20

\bibitem[{Appenzeller \& Mundt(1989)}]{appenzeller1989t}
Appenzeller, I. \& Mundt, R. 1989, The Astronomy and Astrophysics Review, 1,
  291

\bibitem[{Barentsen {et~al.}(2011)Barentsen, Vink, Drew, Greimel, Wright,
  Drake, Martin, Valdivielso, \& Corradi}]{Barentsen2011}
Barentsen, G., Vink, J.~S., Drew, J.~E., {et~al.} 2011, Monthly Notices of the
  Royal Astronomical Society, 415, 103

\bibitem[{Beccari {et~al.}(2015)Beccari, Marchi, Panagia, Valenti, Carraro,
  Romaniello, Zoccali, \& Weidner}]{Beccari2015}
Beccari, G., Marchi, G.~D., Panagia, N., {et~al.} 2015, Astronomy {\&}
  Astrophysics, 574, A44

\bibitem[{{Beccari} {et~al.}(2010){Beccari}, {Spezzi}, {Young}, {De Marchi},
  {Paresce}, {Sirianni}, {Andersen}, \& {SOC}}]{2010AAS...21546329B}
{Beccari}, G., {Spezzi}, L., {Young}, E., {et~al.} 2010, in American
  Astronomical Society Meeting Abstracts, Vol. 215, American Astronomical
  Society Meeting Abstracts \#215, 463.29

\bibitem[{Bessell {et~al.}(1998)Bessell, Castelli, \& Plez}]{bessell1998model}
Bessell, M., Castelli, F., \& Plez, B. 1998, Astronomy and astrophysics, 333,
  231

\bibitem[{Biazzo {et~al.}(2019)Biazzo, Beccari, De~Marchi, \&
  Panagia}]{biazzo2019photometric}
Biazzo, K., Beccari, G., De~Marchi, G., \& Panagia, N. 2019, The Astrophysical
  Journal, 875, 51

\bibitem[{Calvet {et~al.}(2000)Calvet, Hartmann, \& Strom}]{calvet}
Calvet, N., Hartmann, L., \& Strom, E. 2000 (University of Arizona Press), 377

\bibitem[{Calvet {et~al.}(2004)Calvet, Muzerolle, Briceno, Hern{\'a}ndez,
  Hartmann, Saucedo, \& Gordon}]{calvet2004mass}
Calvet, N., Muzerolle, J., Briceno, C., {et~al.} 2004, The Astronomical
  Journal, 128, 1294

\bibitem[{Cardelli {et~al.}(1989)Cardelli, Clayton, \&
  Mathis}]{cardelli1989relationship}
Cardelli, J.~A., Clayton, G.~C., \& Mathis, J.~S. 1989, The Astrophysical
  Journal, 345, 245

\bibitem[{Chen {et~al.}(2015)Chen, Bressan, Girardi, Marigo, Kong, \&
  Lanza}]{chen2015parsec}
Chen, Y., Bressan, A., Girardi, L., {et~al.} 2015, Monthly Notices of the Royal
  Astronomical Society, 452, 1068

\bibitem[{Chiosi {et~al.}(2006)Chiosi, Vallenari, Held, Rizzi, \&
  Moretti}]{chiosi2006age}
Chiosi, E., Vallenari, A., Held, E., Rizzi, L., \& Moretti, A. 2006, Astronomy
  \& Astrophysics, 452, 179

\bibitem[{Clarke \& Pringle(2006)}]{clarke2006m}
Clarke, C. \& Pringle, J. 2006, Monthly Notices of the Royal Astronomical
  Society: Letters, 370, L10

\bibitem[{Dahm(2008)}]{Dahm2008}
Dahm, S.~E. 2008, The Astronomical Journal, 136, 521

\bibitem[{De~Marchi {et~al.}(2013{\natexlab{a}})De~Marchi, Beccari, \&
  Panagia}]{DeMarchi2013}
De~Marchi, G., Beccari, G., \& Panagia, N. 2013{\natexlab{a}}, The
  Astrophysical Journal, 775, 68

\bibitem[{{De Marchi} {et~al.}(2023){De Marchi}, {Giardino}, {Panagia}, {Alves
  de Oliveira}, {Beck}, {Biazzo}, {Brandl}, {Chu}, {Fahrion}, {Habel},
  {Hirschauer}, {Jerabkova}, {Jones}, {Keyes}, {Lenkic}, {Maiolino}, {Meixner},
  {Nally}, {Muzerolle}, {Nayak}, {Pontoppidan}, {Robberto}, {Rogers}, {Sabbi},
  {Sargent}, {Soderblom}, {Zeidler}, \& {NIRSpec Instrument Science
  Team}}]{demarchi2023}
{De Marchi}, G., {Giardino}, G., {Panagia}, N., {et~al.} 2023, in American
  Astronomical Society Meeting Abstracts, Vol.~55, American Astronomical
  Society Meeting Abstracts, 444.07

\bibitem[{De~Marchi {et~al.}(2017)De~Marchi, Panagia, \& Beccari}]{Marchi_2017}
De~Marchi, G., Panagia, N., \& Beccari, G. 2017, The Astrophysical Journal,
  846, 110

\bibitem[{De~Marchi {et~al.}(2013{\natexlab{b}})De~Marchi, Panagia, Guarcello,
  \& Bonito}]{De_Marchi_2013}
De~Marchi, G., Panagia, N., Guarcello, M.~G., \& Bonito, R. 2013{\natexlab{b}},
  Monthly Notices of the Royal Astronomical Society, 435, 3058

\bibitem[{De~Marchi {et~al.}(2010)De~Marchi, Panagia, \&
  Romaniello}]{De_Marchi_2010}
De~Marchi, G., Panagia, N., \& Romaniello, M. 2010, The Astrophysical Journal,
  715, 1

\bibitem[{De~Marchi {et~al.}(2011{\natexlab{a}})De~Marchi, Panagia, Romaniello,
  Sabbi, Sirianni, Moroni, \& Degl{\textquotesingle}Innocenti}]{De_Marchi_2011}
De~Marchi, G., Panagia, N., Romaniello, M., {et~al.} 2011{\natexlab{a}}, The
  Astrophysical Journal, 740, 11

\bibitem[{De~Marchi {et~al.}(2011{\natexlab{b}})De~Marchi, Paresce, Panagia,
  Beccari, Spezzi, Sirianni, Andersen, Mutchler, Balick, Dopita, Frogel,
  Whitmore, Bond, Calzetti, Carollo, Disney, Hall, Holtzman, Kimble, McCarthy,
  O{\textquotesingle}Connell, Saha, Silk, Trauger, Walker, Windhorst, \&
  Young}]{DeMarchi2011}
De~Marchi, G., Paresce, F., Panagia, N., {et~al.} 2011{\natexlab{b}}, The
  Astrophysical Journal, 739, 27

\bibitem[{Drew {et~al.}(2005)Drew, Greimel, Irwin, Aungwerojwit, Barlow,
  Corradi, Drake, Gänsicke, Groot, Hales, Hopewell, Irwin, Knigge, Leisy,
  Lennon, Mampaso, Masheder, Matsuura, Morales-Rueda, Morris, Parker,
  Phillipps, Rodriguez-Gil, Roelofs, Skillen, Sokoloski, Steeghs, Unruh,
  Viironen, Vink, Walton, Witham, Wright, Zijlstra, \&
  Zurita}]{10.1111/j.1365-2966.2005.09330.x}
Drew, J.~E., Greimel, R., Irwin, M.~J., {et~al.} 2005, Monthly Notices of the
  Royal Astronomical Society, 362, 753

\bibitem[{Evans {et~al.}(2006)Evans, Lennon, Smartt, \&
  Trundle}]{evansetal2006}
Evans, C., Lennon, D., Smartt, S., \& Trundle, C. 2006, Astronomy \&
  Astrophysics, 456, 623

\bibitem[{Gennaro \& et~al.(2018)}]{wfc3book}
Gennaro, M. \& et~al. 2018, {WFC3 Data Handbook}

\bibitem[{Gouliermis {et~al.}(2007)Gouliermis, Henning, Brandner, Dolphin,
  Rosa, \& Brandl}]{gouliermisetal2007}
Gouliermis, D.~A., Henning, T., Brandner, W., {et~al.} 2007, The Astrophysical
  Journal Letters, 665, L27

\bibitem[{Gullbring {et~al.}(1998)Gullbring, Hartmann, Briceno, \&
  Calvet}]{gullbring1998disk}
Gullbring, E., Hartmann, L., Briceno, C., \& Calvet, N. 1998, The Astrophysical
  Journal, 492, 323

\bibitem[{{Hartmann} {et~al.}(1998){Hartmann}, {Calvet}, {Gullbring}, \&
  {D'Alessio}}]{1998ApJ...495..385H}
{Hartmann}, L., {Calvet}, N., {Gullbring}, E., \& {D'Alessio}, P. 1998, \apj,
  495, 385

\bibitem[{Hartmann {et~al.}(2016)Hartmann, Herczeg, \&
  Calvet}]{doi:10.1146/annurev-astro-081915-023347}
Hartmann, L., Herczeg, G., \& Calvet, N. 2016, Annual Review of Astronomy and
  Astrophysics, 54, 135

\bibitem[{Hennekemper {et~al.}(2008)Hennekemper, Gouliermis, Henning, Brandner,
  \& Dolphin}]{hennekemperetal2008}
Hennekemper, E., Gouliermis, D.~A., Henning, T., Brandner, W., \& Dolphin,
  A.~E. 2008, The Astrophysical Journal, 672, 914

\bibitem[{Joy(1945)}]{joy1945t}
Joy, A.~H. 1945, The Astrophysical Journal, 102, 168

\bibitem[{Kroupa(2001)}]{kroupa2001variation}
Kroupa, P. 2001, Monthly Notices of the Royal Astronomical Society, 322, 231

\bibitem[{Lee {et~al.}(2005)Lee, Rolleston, Dufton, \& Ryans}]{lee2005chemical}
Lee, J.-K., Rolleston, W., Dufton, P., \& Ryans, R. 2005, Astronomy \&
  Astrophysics, 429, 1025

\bibitem[{Lucas \& et~al.(2021)}]{acsbook}
Lucas, R.~A. \& et~al. 2021, {ACS Data Handbook for Cycle 29}

\bibitem[{{Martayan, C.} {et~al.}(2010){Martayan, C.}, {Baade, D.}, \&
  {Fabregat, J.}}]{refId0}
{Martayan, C.}, {Baade, D.}, \& {Fabregat, J.} 2010, A\&A, 509, A11

\bibitem[{Massey {et~al.}(1989)Massey, Parker, \& Garmany}]{masseyetal1989}
Massey, P., Parker, J.~W., \& Garmany, C.~D. 1989, The Astronomical Journal,
  98, 1305

\bibitem[{Muzerolle {et~al.}(2003)Muzerolle, Hillenbrand, Calvet, Briceno, \&
  Hartmann}]{muzerolle2003accretion}
Muzerolle, J., Hillenbrand, L., Calvet, N., Briceno, C., \& Hartmann, L. 2003,
  The Astrophysical Journal, 592, 266

\bibitem[{Muzerolle {et~al.}(2005)Muzerolle, Luhman, Briceno, Hartmann, \&
  Calvet}]{muzerolle2005measuring}
Muzerolle, J., Luhman, K.~L., Briceno, C., Hartmann, L., \& Calvet, N. 2005,
  The Astrophysical Journal, 625, 906

\bibitem[{Natta {et~al.}(2004)Natta, Testi, Muzerolle, Randich, Comer{\'o}n, \&
  Persi}]{natta2004accretion}
Natta, A., Testi, L., Muzerolle, J., {et~al.} 2004, Astronomy \& Astrophysics,
  424, 603

\bibitem[{Natta {et~al.}(2006)Natta, Testi, \& Randich}]{natta2006accretion}
Natta, A., Testi, L., \& Randich, S. 2006, Astronomy \& Astrophysics, 452, 245

\bibitem[{Nota {et~al.}(2006)Nota, Sirianni, Sabbi, Tosi, Clampin, Gallagher,
  Meixner, Oey, Pasquali, Smith, {et~al.}}]{notaetal2006}
Nota, A., Sirianni, M., Sabbi, E., {et~al.} 2006, The Astrophysical Journal
  Letters, 640, L29

\bibitem[{Noël {et~al.}(2007)Noël, Gallart, Costa, \& Méndez}]{Noel_2007}
Noël, N. E.~D., Gallart, C., Costa, E., \& Méndez, R.~A. 2007, The
  Astronomical Journal, 133, 2037

\bibitem[{{Perez} \& {Grady}(1997)}]{1997SSRv...82..407P}
{Perez}, M.~R. \& {Grady}, C.~A. 1997, \ssr, 82, 407

\bibitem[{P{\'e}rez-Montero \& D{\'\i}az(2005)}]{perez2005comparative}
P{\'e}rez-Montero, E. \& D{\'\i}az, A.~I. 2005, Monthly Notices of the Royal
  Astronomical Society, 361, 1063

\bibitem[{Piatti {et~al.}(2007)Piatti, Sarajedini, Geisler, Clark, \&
  Seguel}]{piatti2007young}
Piatti, A.~E., Sarajedini, A., Geisler, D., Clark, D., \& Seguel, J. 2007,
  Monthly Notices of the Royal Astronomical Society, 377, 300

\bibitem[{Rivinius {et~al.}(2013)Rivinius, Carciofi, \&
  Martayan}]{Rivinius_2013}
Rivinius, T., Carciofi, A.~C., \& Martayan, C. 2013, The Astronomy and
  Astrophysics Review, 21

\bibitem[{Rolleston {et~al.}(1999)Rolleston, Dufton, McErlean, \&
  Venn}]{rolleston1999chemical}
Rolleston, W., Dufton, P., McErlean, N., \& Venn, K. 1999, Astronomy and
  Astrophysics, 348, 728

\bibitem[{Romaniello(1998)}]{phd}
Romaniello, M. 1998, PhD thesis, Scuola Normale Superiore, Pisa, Italy

\bibitem[{Russell \& Dopita(1992)}]{russell1992abundances}
Russell, S.~C. \& Dopita, M.~A. 1992, The Astrophysical Journal, 384, 508

\bibitem[{Sabbi {et~al.}(2011)Sabbi, Nota, Tosi, Smith, Gallagher, \&
  Cignoni}]{Sabbi}
Sabbi, E., Nota, A., Tosi, M., {et~al.} 2011, The Astrophysical Journal, 739,
  15

\bibitem[{Sabbi {et~al.}(2006)Sabbi, Sirianni, Nota, Tosi, Gallagher, Meixner,
  Oey, Walterbos, Pasquali, Smith, {et~al.}}]{sabbietal2007}
Sabbi, E., Sirianni, M., Nota, A., {et~al.} 2006, The Astronomical Journal,
  133, 44

\bibitem[{{Savage} \& {Mathis}(1979)}]{1979ARA&A..17...73S}
{Savage}, B.~D. \& {Mathis}, J.~S. 1979, \araa, 17, 73

\bibitem[{Schmalzl {et~al.}(2008)Schmalzl, Gouliermis, Dolphin, \&
  Henning}]{Schmalzl_2008}
Schmalzl, M., Gouliermis, D.~A., Dolphin, A.~E., \& Henning, T. 2008, The
  Astrophysical Journal, 681, 290

\bibitem[{{Shu}(1977)}]{1977ApJ...214..488S}
{Shu}, F.~H. 1977, \apj, 214, 488

\bibitem[{Sicilia-Aguilar {et~al.}(2005)Sicilia-Aguilar, Hartmann,
  Hern{\'a}ndez, Briceno, \& Calvet}]{sicilia2005cepheus}
Sicilia-Aguilar, A., Hartmann, L.~W., Hern{\'a}ndez, J., Briceno, C., \&
  Calvet, N. 2005, The Astronomical Journal, 130, 188

\bibitem[{Spezzi {et~al.}(2012)Spezzi, De~Marchi, Panagia, Sicilia-Aguilar, \&
  Ercolano}]{10.1111/j.1365-2966.2011.20130.x}
Spezzi, L., De~Marchi, G., Panagia, N., Sicilia-Aguilar, A., \& Ercolano, B.
  2012, Monthly Notices of the Royal Astronomical Society, 421, 78

\bibitem[{Tang {et~al.}(2014)Tang, Bressan, Rosenfield, Slemer, Marigo,
  Girardi, \& Bianchi}]{tang2014new}
Tang, J., Bressan, A., Rosenfield, P., {et~al.} 2014, Monthly Notices of the
  Royal Astronomical Society, 445, 4287

\bibitem[{Tognelli {et~al.}(2011)Tognelli, Moroni, \&
  Degl’Innocenti}]{tognelli2011pisa}
Tognelli, E., Moroni, P.~P., \& Degl’Innocenti, S. 2011, Astronomy \&
  Astrophysics, 533, A109

\bibitem[{Vieira {et~al.}(2021)Vieira, Garc\'ia-Varela, Sabogal, R\'imulo, \&
  Hern\'andez}]{10.1093/mnras/stab1481}
Vieira, K., Garc\'ia-Varela, A., Sabogal, B., R\'imulo, L.~R., \& Hern\'andez,
  J. 2021, Monthly Notices of the Royal Astronomical Society, 505, 5567

\bibitem[{Waters \& Waelkens(1998)}]{waters1998herbig}
Waters, L. \& Waelkens, C. 1998, Annual Review of Astronomy and Astrophysics,
  36, 233

\bibitem[{White \& Basri(2003)}]{white2003very}
White, R.~J. \& Basri, G. 2003, The Astrophysical Journal, 582, 1109

\bibitem[{White \& Hillenbrand(2004)}]{white2004evolutionary}
White, R.~J. \& Hillenbrand, L.~A. 2004, The Astrophysical Journal, 616, 998

\bibitem[{Witham {et~al.}(2008)Witham, Knigge, Drew, Greimel, Steeghs,
  Gänsicke, Groot, \& Mampaso}]{10.1111/j.1365-2966.2007.12774.x}
Witham, A.~R., Knigge, C., Drew, J.~E., {et~al.} 2008, Monthly Notices of the
  Royal Astronomical Society, 384, 1277

\end{thebibliography}
%

\end{document}